\documentclass[fleqn,usenatbib,useAMS]{mnras}

\usepackage{graphicx}	
\usepackage{amsmath}	
\usepackage{amssymb}	
\usepackage{multicol}        
\usepackage{bm}		
\usepackage{pdflscape}	

\usepackage[T1]{fontenc}
\usepackage{ae,aecompl}

\usepackage{newtxtext,newtxmath}

\date{Last updated 2020 June 10; in original form 2013 September 5}

\pubyear{2022}

\date{Last updated 2020 June 10; in original form 2013 September 5}

\pubyear{2022}

\title[Mixing in pulsar wind nebulae]{Material mixing in pulsar wind nebulae of massive runaway stars \\ }

\author[D. M.-A.~Meyer et al.]
       {D. M.-A.~Meyer\thanks{E-mail: meyer@ice.csic.es }$^{1}$ and 
       D.~F.~Torres\thanks{E-mail: dtorres@ice.csic.es}$^{1,2,3}$
       \\
       $^{1}$ Institute of Space Sciences (ICE, CSIC), Campus UAB, Carrer 
              de Can Magrans s/n, 08193 Barcelona, Spain \\       
       $^{2}$ Institut d'Estudis Espacials de Catalunya (IEEC), 
              08860 Barcelona, Spain \\ 
       $^{3}$ Instituci\'o Catalana de Recerca i Estudis Avan\c{c}ats (ICREA), Barcelona, Spain \\        
       }

\begin{document}
\label{firstpage}
\pagerange{\pageref{firstpage}--\pageref{lastpage}}
\maketitle

\begin{abstract}
%
   In this study we  quantitatively examine the manner 
   pulsar wind, supernova ejecta and defunct stellar wind materials distribute and melt 
   together into plerions. 
   We performed 2.5D MHD simulations of the entire evolution 
   of their stellar surroundings and different scenarios are explored, whether the star dies as 
   a red supergiant and Wolf-Rayet supernova progenitors, and whether it moved with velocity 
   $20\, \rm km\, \rm s^{-1}$ or $40\, \rm km\, \rm s^{-1}$ through the ISM. 
   Within the post-explosion, early $10\, \rm kyr$, the H-burning-products rich  
   red supergiant wind only mixes by $\le 20\%$, due to its dense circumstellar medium filling 
   the progenitor's bow shock trail, still unaffected by the supernova blastwave. Wolf-Rayet 
   materials, enhanced in C, N, O elements, distribute circularly for the 
   $35\, \rm M_\odot$ star moving at $20\, \rm km\, \rm s^{-1}$ and oblongly at higher 
   velocities, mixing efficiently up to $80\%$.
   Supernova ejecta, \textcolor{black}{filled with Mg, Si, Ca, Ti and Fe,} remain 
   spherical for longer times at $20\, \rm km\, \rm s^{-1}$ but form 
   complex patterns at higher progenitor speeds due to earlier interaction with the bow shock, 
   in which they mix more efficiently. 
   The pulsar wind mixing is more efficient for Wolf-Rayet ($25\%$) than red supergiant 
   progenitors ($20\%$).
   This work reveals that the past evolution of massive stars and their circumstellar 
   environments critically shapes the internal distribution of chemical elements on 
   plerionic supernova remnants, and, therefore, governs the origin 
   of the various emission mechanisms at work therein. This is essential for 
   interpreting multi-frequency observations of atomic and molecular 
   spectral lines, such as in optical, infrared, and soft X-rays. 
\end{abstract}



\section{Introduction}
\label{intro}

Massive stars form in dense, opaque molecular clouds which contracts under 
their own gravity. 
Once local gravitational collapse takes place and form 
a stellar embryo, the conservation of momentum organises the infalling material 
as an accretion disc, which fragments. Its debris feed the growing protostar, 
inducing luminous flares and forming companions 
\citep{meyer_mnras_482_2019,2024arXiv240519905M}.
When the gravity of the massive protostar is sufficiently important, fusion 
nuclear reactions ignite in its core, expelling strong stellar winds and 
radiation from its surface, which dissipate the molecular environment 
and shape it as wind bubbles \citep{weaver_apj_218_1977}. 
If the star 
moves fast enough, the bubble is distorted as an arc-like nebula called bow 
shock \citep{gull_apj_230_1979,wilkin_459_apj_1996}. The circumstellar 
medium of massive stars continues reflecting stellar feedback throughout
their entire evolution, e.g. as a red supergiant or a Wolf-Rayet object, 
see \citet{avedisova_saj_15_1972,smith_mnras_211_1984,noriegacrespo_aj_114_1997}. 
Some high-mass stars, at the end of their lives, explode as a core-collapse 
supernova, releasing energy and mass into their pre-shaped surroundings. 
The meeting of the supernova blastwave with the wind nebula produces a supernova 
remnant, and additionally, when a compact object such as a rotating magnetized 
neutron star form as the ultimate rest of the massive star, a relativistic wind 
power the so-called plerionic supernova remnant \citep{weiler_aa_70_1978, 
caswell_mnras_187_1979, weiler_aa_90_1980, Gaensler_Slane_2006ARA&A..44...17G,
Kargaltsev_etal_2017JPlPh..83e6301K}.

In addition to a growing accumulation of multi-wavelengths observations, see e.g.
\citep{Hester_2008ARA&A..46..127H, behler_RPPh_2014, bock_aj_1998, Kargaltsev_apss_308_2007, 
Kargaltsev_apjs_201_2012,2017JPlPh..83e6301K, frail_apj_480_1997,hess_aa_612_2018,
popov_2019, 2019A&A...627A.100H,acero_aph_2023,turner_mnras_531_2024}, 
%
%
the understandings of pulsar wind nebulae and pulsar-powered supernova remnants 
is inseparable from the development of numerical simulations.
%
Series of 1-dimensional works \citep{kennel_apj_283_1984,coroniti_apj_349_1990,
begelman_apj_397_1992,begelman_apj_493_1998}, 2-dimensional models \citet{komissarov_mnras_344L_2003,komissarov_mnras_349_2004, 
swaluw_aa_397_2003,swaluw_aa_420_2004, 
komissarov_mnras_367_2006,Del_Zanna_etal_2006A&A...453..621D, 
Camus_etal_2009MNRAS.400.1241C, komissarov_mnras_414_2011, 
Olmi_etal_2014MNRAS.438.1518O} , 3-dimensional studies 
\citep{2013MNRAS.431L..48P,Porth_etal_2014MNRAS.438..278P, 
Olmi_etla_2016JPlPh..82f6301O} and applications to the plerions 
Geminga and Vela have been performed \citep{Hester_2008ARA&A..46..127H, 
behler_RPPh_2014, bock_aj_1998, popov_2019, 2019A&A...627A.100H}, deepening 
our comprehension of the internal functioning of pulsar wind nebulae, 
from the magnetosphere to the pc-scale environment. 
The reverberation phase of the pulsar wind termination shock 
\cite{Bandiera_mnras_499_2020,Bandiera_mnras_165_2023,2023_mnras_2839_2023},
and the motion of the pulsar inside of the supernova remnant 
\citep{temim_apj_808_2015,temim_apj_851_2017,kolb_apj_844_2017,temim_apj_932_2022} 
and beyond, into the ISM \citep{Bucciantini_aa_375_2001,bucciantini_mnras_478_2018,
Toropina_etal_2019MNRAS.484.1475T}, have been particularly focused upon.

\textcolor{black}{
Until now, numerical simulations of pulsar wind nebulae have not considered the detailed chemical composition 
of the various materials involved in the problem, which correspond to the wind blown during the main-sequence, 
red supergiant, and Wolf-Rayet evolutionary phases of the progenitor, as well as to the supernova ejecta released 
after the explosion \citep{maeder_araa_38_2000}. 
%
Mixing of the dominant chemical species of each component will happen at the unstable discontinuities 
separating the various regions.
%
Massive stars spend most of their lives in a long main-sequence phase, during which both the surface chemical 
composition and the composition of the accelerated stellar wind are primarily made up of hydrogen (H) 
As main-sequence massive stars evolve into the red supergiant phase, the composition of the stellar surface 
continues to be dominated by H-atoms, even though their cores are depleted of H and begin to fuse helium (He). 
Towards the end of the supergiant phase, the surface becomes enriched in He to a level comparable to that of H. 
The winds of red supergiants are enriched in nitrogen (N) and sulfur (S), as evidenced by the emission lines 
observed from the bow shock nebula of the runaway red supergiant IRC-10414 
\citep{2012A&A...537A..35C,meyer_2014a,Gvaramadze_2013,meyer_mnras_506_2021}.
}

\textcolor{black}{
If the stellar mass is sufficiently high, its radius, surface, and wind properties evolve again as the star 
enters the Wolf-Rayet evolutionary phase \citep{maeder_araa_38_2000}. Throughout this phase, carbon (C), neon (Ne), 
magnesium (Mg), followed by oxygen (O) and finally silicon (Si), are produced by nuclear fusion, 
generating an iron (Fe) core that ignites the final mechanism for a core-collapse supernova explosion 
\citep{smartt_araa_47_2009,janka_arnps_66_2016}. The winds of Wolf-Rayet stars are enriched in He, 
C, N and O \citep{Hamann2006} and, in their turn, enrich their circumstellar medium accordingly 
\citep{maeder_2009,berlanas_aa_620_2018,schulreich_aa_680_2023,gomez_gomzalez_mnras_509_2022}. 
The respective proportions of the chemical yields constituting the winds of massive stars are further 
affected by their zero-age main-sequence stellar rotation and the metallicity of their native 
environment\citep{mokiem_aa_465_2007,brott_aa_530_2011b,vink_mnras_504_2021,marcolino_aa_690_2024}. 
Notably, the most massive stars with initial masses $\ge 100\, \rm M_{\odot}$ can process heavy nuclei in their 
cores via the sodium-neon (Na-Ne) and magnesium-aluminum (Mg-Al) cycles, releasing this aluminum (Al) excess 
into their winds and replenishing the interstellar medium with it \citep{dearborn_apj_277_1984}. 
}

\textcolor{black}{
The supernova explosion releases all these chemical elements into the circumstellar medium  
\citep{1975ApJ...195..715M,martizzi_mnras_405_2015,orlando_aa_622_2019, orlando_aa_645_2021, orland_aa_636_2020, 
2022arXiv220201643O, orlando_aa_666_2022} when the blast wave collides with the wind nebula formed 
throughout the star's life, resulting in a supernova remnant that is eventually powered by a pulsar 
wind \citep{weiler_aa_70_1978, caswell_mnras_187_1979, weiler_aa_90_1980,reynolds_apj_278_1984,
2017hsn_book_2159S}. 
The distribution of chemical species within plerions is therefore governed 
by the local conditions of the interstellar medium, the stellar evolution history of the progenitor, the 
properties of the explosion and the characteristics of the pulsar. 
These interstellar, circumstellar 
and ejecta chemical elements, ruling the local cooling and heating processes of the gas 
by atomic and molecular emission \citep{wolfire_apj_587_2003,tesileanu_aa_488_2008,wiersma_mnras_393_2009}, 
which generate X-rays, radio, infrared and optical emission, mostly produced by H, He, O, Fe and 
a plethora of other atoms, ions and molecules\citep{whiteoak_aas_118_1996,reach_aj_131_2006,seok_apj_779_2013}. 
An example of such supernova remnant is the Cygnus Loop and its ejecta distribution of heavy 
elements \citep{uchida_pasj_61_2009}. 
}

\textcolor{black}{
The chemical stellar wind history is therefore a factor that 
}
can not be excluded when modelling the long-term ($>10\, \rm kyr$) evolution of 
pulsar wind nebulae to understand \textcolor{black}{their emission. }
Numerical simulations provided so far two examples. First, that 
of a runaway supernova progenitor which wind-ISM produces a stellar wind bow shock 
inducing asymmetries in the pulsar wind nebula \citep{meyer_mnras_515_2022}. 
Secondly, the case of a static massive star having the termination shock of 
its stellar wind bubble elongated by the organised magnetic field of the ISM. 
The resulting pulsar wind nebula develops in its turn asymmetries reflecting 
that of its pre-supernova circumstellar medium \citep{meyer_527_mnras_2024}. 
%
%
We hereby continue this numerical effort, by tackling the question of the mixing of 
the several kind of plasma participating in the constitution of the plerions.

This study is organized as follows. In Section \ref{method}, we review the simulation 
strategy, the different utilised methods and initial conditions used to carry out the 
magneto-hydrodynamical (MHD) simulations of the evolving close medium of a series of runaway 
high-mass stars, from the onset of their main-sequence to their supernova 
remnant phase. 
Section \ref{results} provides a detailed description of the internal distribution of 
materials (stellar winds, supernova ejecta, pulsar wind) into the pulsar wind nebula 
produced by the moving progenitors. We particularly study the mixing at work therein. 
Section \ref{discussion} further discusses the results, their limitation and scope. 
Finally, we draw our conclusions in Section \ref{conclusion}.


\begin{table}
	\centering
	\caption{
	List of numerical models. All simulations assume a rotating massive progenitor of zero-age main-sequence 
	mass $M_{\star}$ (in $\rm M_{\odot}$) at solar metallicity and moving with velocity $v_{\star}$ through 
	the warm phase of the Galactic plane. 
    The table indicate the stellar evolution history in each model, from the MS (main-sequence) phase 
    to the final SN (supernova) explosion and ultimate PWN (pulsar wind), through the RSG (red supergiant) 
    and Wolf-Rayet stages. 
	}
	\begin{tabular}{lcccr}
	\hline
	${\rm {Model}}$       &    Evolution history     \\ 
	\hline   
	$\rm PWN-20Mo-v20-mix$ &  $\rm MS \rightarrow \rm RSG \rightarrow \rm SN \rightarrow \rm PWN$           \\
	$\rm PWN-20Mo-v40-mix$ &  $\rm MS \rightarrow \rm RSG \rightarrow \rm SN \rightarrow \rm PWN$           \\
	$\rm PWN-35Mo-v20-mix$ &  $\rm MS \rightarrow \rm RSG \rightarrow \rm WR \rightarrow \rm SN \rightarrow \rm PWN$ \\
	$\rm PWN-35Mo-v40-mix$ &  $\rm MS \rightarrow \rm RSG \rightarrow \rm WR \rightarrow \rm SN \rightarrow \rm PWN$ \\
	\hline    
    \footnotesize 
	\end{tabular}
\label{tab:table1}
\end{table}

\begin{figure*}
        \centering
        \includegraphics[width=1.8\columnwidth]{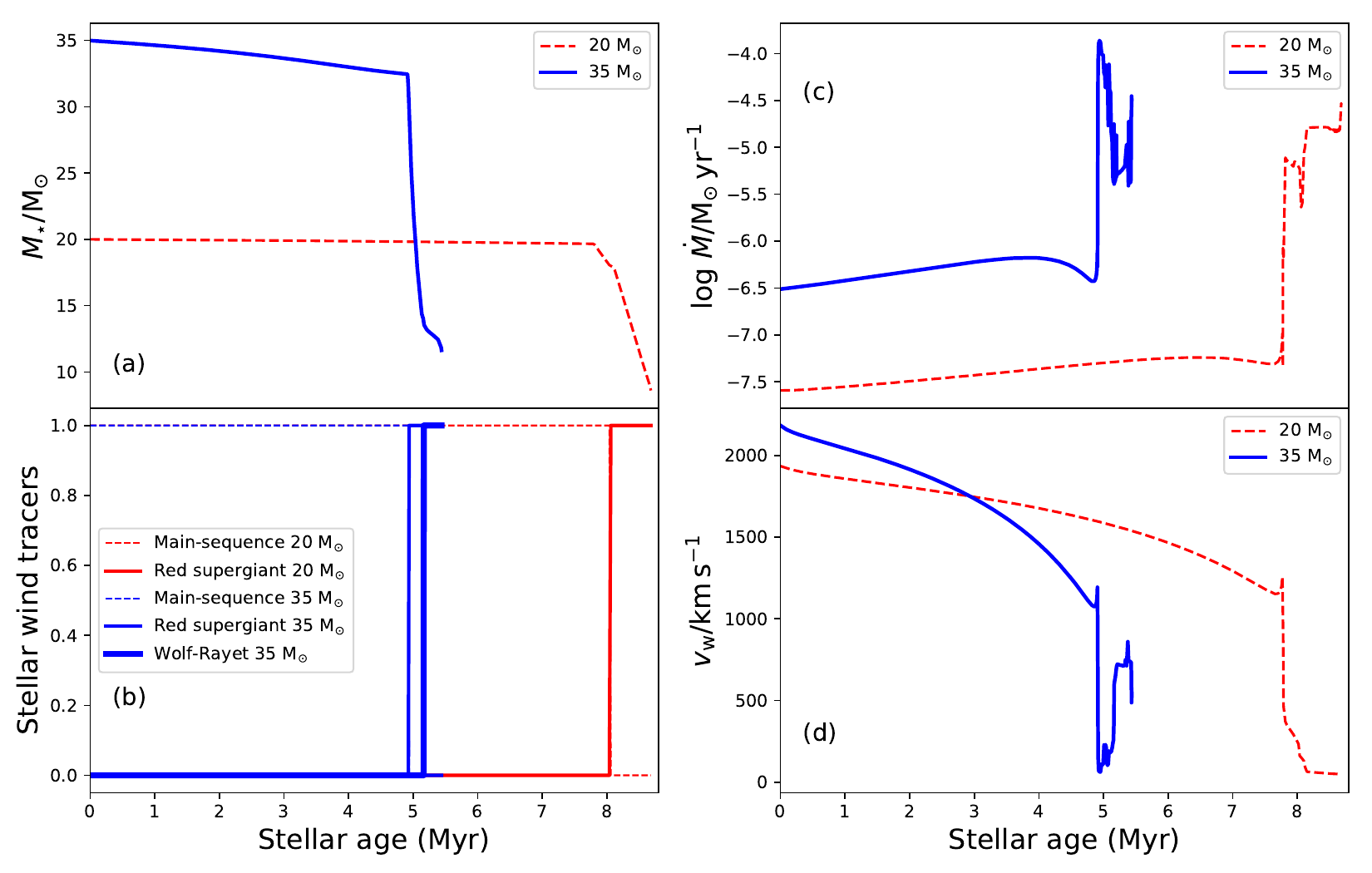}  \\      
        \caption{
        Time evolution (in $\rm Myr$) of the supernova progenitors of 
        zero-age main-sequence $20\, \rm M_{\odot}$ (dotted red line) 
        and $35\, \rm M_{\odot}$ (solid blue line) which we use in our work. 
        The figures show the mass of the stars $M_{\star}$ (panel a, in $\rm M_{\odot}$),
        the values of the stellar wind tracers (panel b), 
        their mass-loss rate $\dot{M}$ (panel c, in $\rm M_{\odot}\, \rm yr^{-1}$).         
        and the wind velocity $v_{\rm w}$ (panel d, in $\rm km\ \rm s^{-1}$). 
        }
        \label{fig:plot_star_properties}  
\end{figure*}

\section{Method}
\label{method}


\subsection{Simulation strategy}
\label{method_start}

%
The global modelling of young to middle-age pulsar wind nebulae is worked in a 
three-steps fashion, and has a particular set of advantages and caveats.

First, the circumstellar medium generated by stellar 
wind-ISM interaction of the moving massive star is simulated from its zero-age 
main-sequence time to its pre-supernova time. The wind-ISM interaction of the 
progenitor is calculated in the reference frame of the moving star, and serves 
as the initial condition for the calculation of the interaction between the 
blastwave and the circumstellar medium. 
Secondly, the supernova blastwave, with released ejecta corresponding to the progenitor 
mass at the moment of the explosion, minus the mass of a canonical neutron star, 
is injected into freely-expanding last stellar 
wind and advanced in a 1-dimensional, high-resolution fashion during 50 years, 
which is sufficient to simulate the structure of the region of swept-up stellar 
gas and supernova material that progresses trough the wind. 
Last, this 1D solution is mapped onto the 2.5-dimensional circumstellar medium, 
i.e. the density, velocity and pressure gas structure, described 
by 2 dimensions for the scalar quantities plus a toroidal component for the vectors,
at the moment of the explosion, and a pulsar wind is injected at the location 
of the supernova. 
The whole evolution is then continued within the 2.5-dimensional 
approach.

The evolution of the chemical yields into supernova remnants is a profound 
question to investigate. The main-sequence, red supergiant and Wolf-Rayet 
stellar winds are tracked using a series of passive scalar tracers which 
values are time-dependently set to 1 throughout the corresponding stellar 
evolutionary phases and to 0 elsewhen. 
Similarly, a tracer gives access to the distribution of the supernova ejecta 
and another permits to follow the expanding of the pulsar wind into the 
supernova remnant \citet{meyer_527_mnras_2024,meyer_aa_687_2024}. 
The passive scalar fields are evolved throughout the supernova remnants via the 
 advection equations, 
\begin{equation}
	\frac{\partial (\rho Q_{\rm MS}) }{\partial t } + \vec{ \nabla } \cdot ( \vec{v} \rho Q_{\rm MS}) = 0,
\label{eq:tracer1}
\end{equation}
\begin{equation}
	\frac{\partial (\rho Q_{\rm RSG}) }{\partial t } + \vec{ \nabla } \cdot ( \vec{v} \rho Q_{\rm RSG}) = 0,
\label{eq:tracer2}
\end{equation}
\begin{equation}
	\frac{\partial (\rho Q_{\rm WR}) }{\partial t } + \vec{ \nabla } \cdot ( \vec{v} \rho Q_{\rm WR}) = 0,
\label{eq:tracer3}
\end{equation}
\begin{equation}
	\frac{\partial (\rho Q_{\rm EJ}) }{\partial t } 
	+ \vec{ \nabla } \cdot ( \vec{v} \rho Q_{\rm EJ}) = 0,
\label{eq:tracer4}
\end{equation}
\begin{equation}
	\frac{\partial (\rho Q_{\rm PSR}) }{\partial t } 
	+ \vec{ \nabla } \cdot ( \vec{v} \rho Q_{\rm PSR}) = 0,
\label{eq:tracer5}
\end{equation}
with $Q_{\rm MS}$, $Q_{\rm RSG}$, $Q_{\rm WR}$, $Q_{\rm EJ}$ and $Q_{\rm PSR}$ 
the tracers for the main-sequence, red supergiant, Wolf-Rayet stellar winds, 
supernova ejecta and pulsar wind material, $\rho$ the mass density and $v$ 
the gas velocity, respectively. 
Hence, a 2.5-dimensional plerionic supernova remnants are obtained, 
including a series of 5 tracers allowing us to track the distribution 
of the several species involved into that problem. This approach is similar to that found in
\citet{orland_aa_636_2020,orlando_aa_666_2022}. 
Such approach permits to reach models of high spatial resolution at moderate 
computational costs, which allows us to explore the parameter space of the 
problem.


\begin{figure*}
        \centering
        \includegraphics[width=1.4\columnwidth]{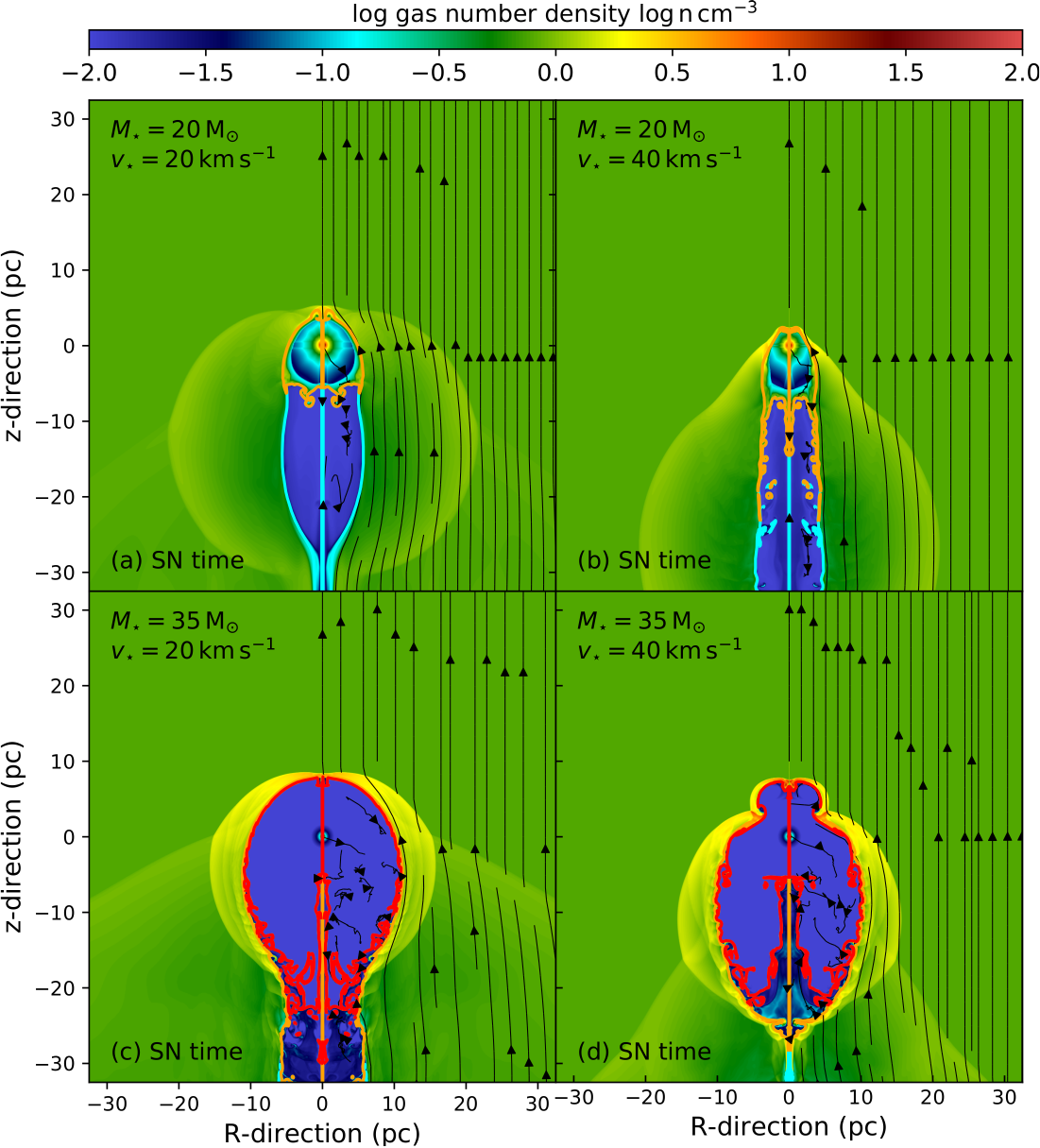}  \\        
        \caption{
        \textcolor{black}{
        Number density fields in our magneto-hydrodynamical simulation of 
        the circumstellar medium of the runaway stars at the supernova time. 
        The figure displays the models for the $20\, \rm M_{\odot}$ (top) 
        and the $35\, \rm M_{\odot}$ (bottom) stars moving at velocities 
        $v_{\star}=20\, \rm km\, \rm s^{-1}$ (left) 
        and $v_{\star}=40\, \rm km\, \rm s^{-1}$ (right). 
        The various contours highlight the region with a $50\%$ contribution of 
        the Wolf-Rayet wind (red), red supergiant 
        wind (orange) and with a $10\%$ contribution of main-sequence material 
        (cyan), respectively. 
        The black arrows in the right-hand parts of the figures are  
        ISM magnetic field lines. 
        }        
        }
        \label{fig:csm_at_psn}  
\end{figure*}

\begin{figure*}
        \centering
        \includegraphics[width=1.4\columnwidth]{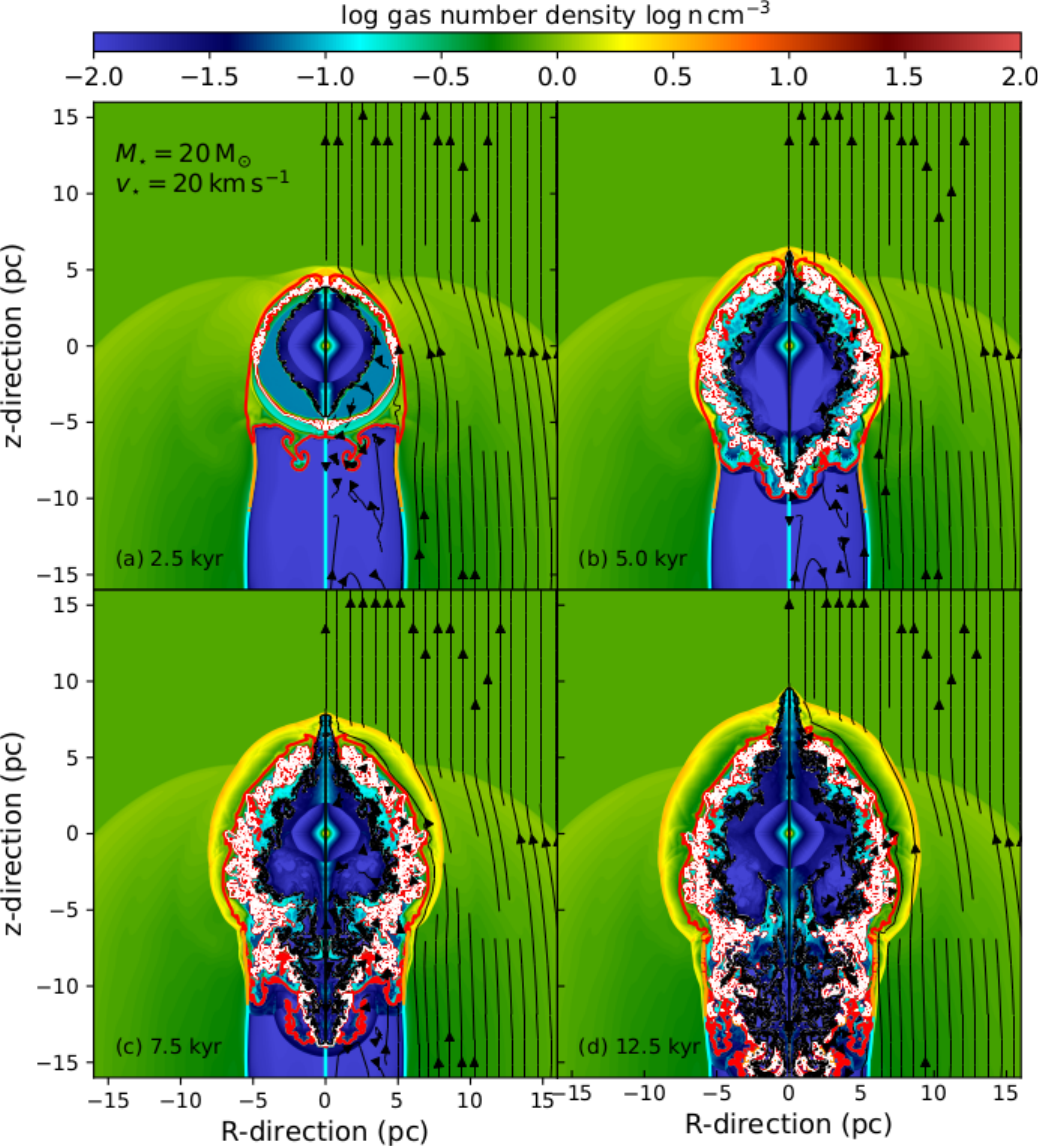}  \\        
        \caption{
        Number density fields in our magneto-hydrodynamical simulation of 
        the supernova remnant of the runaway $20\, \rm M_{\odot}$ star rotating 
        with $\Omega_{\star}/\Omega_{\rm K}=0.1$ and moving with velocity 
        $v_{\star}=20\, \rm km\, \rm s^{-1}$. 
        The evolution of the plerionic supernova remnant is shown at times 
        $2.5$ (a), $5.0$ (b), $7.5$ (c) and $12.5\, \rm kyr$ (d), respectively. 
        The various contours highlight the region with a $50\%$ contribution of 
        pulsar wind nebula (black), 
        supernova ejecta (white), Wolf-Rayet wind (red), red supergiant 
        wind (orange) and with a $10\%$ contribution of main-sequence material 
        (cyan), respectively. 
        The black arrows in the right-hand parts of the figures are  
        ISM magnetic field lines. 
        }
        \label{fig:snr_pwn_2020}  
\end{figure*}

\begin{figure*}
        \centering
        \includegraphics[width=1.4\columnwidth]{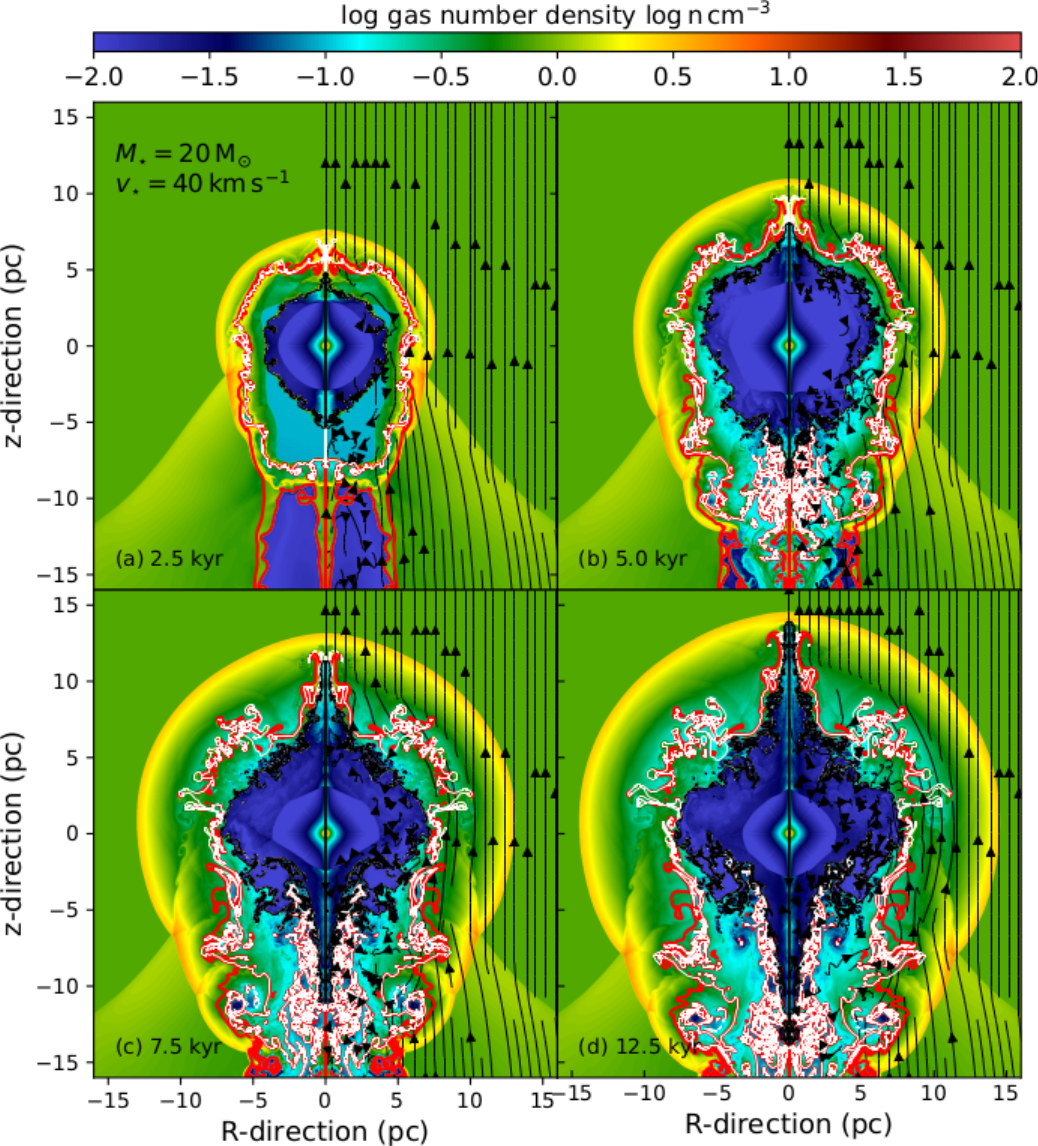}  \\        
        \caption{
        As Fig.~\ref{fig:snr_pwn_2020} for a $20\, \rm M_{\odot}$ progenitor star 
        moving with velocity $v_{\star}=40\, \rm km\, \rm s^{-1}$. 
        }
        \label{fig:snr_pwn_2040}  
\end{figure*}

\begin{figure*}
        \centering
        \includegraphics[width=1.4\columnwidth]{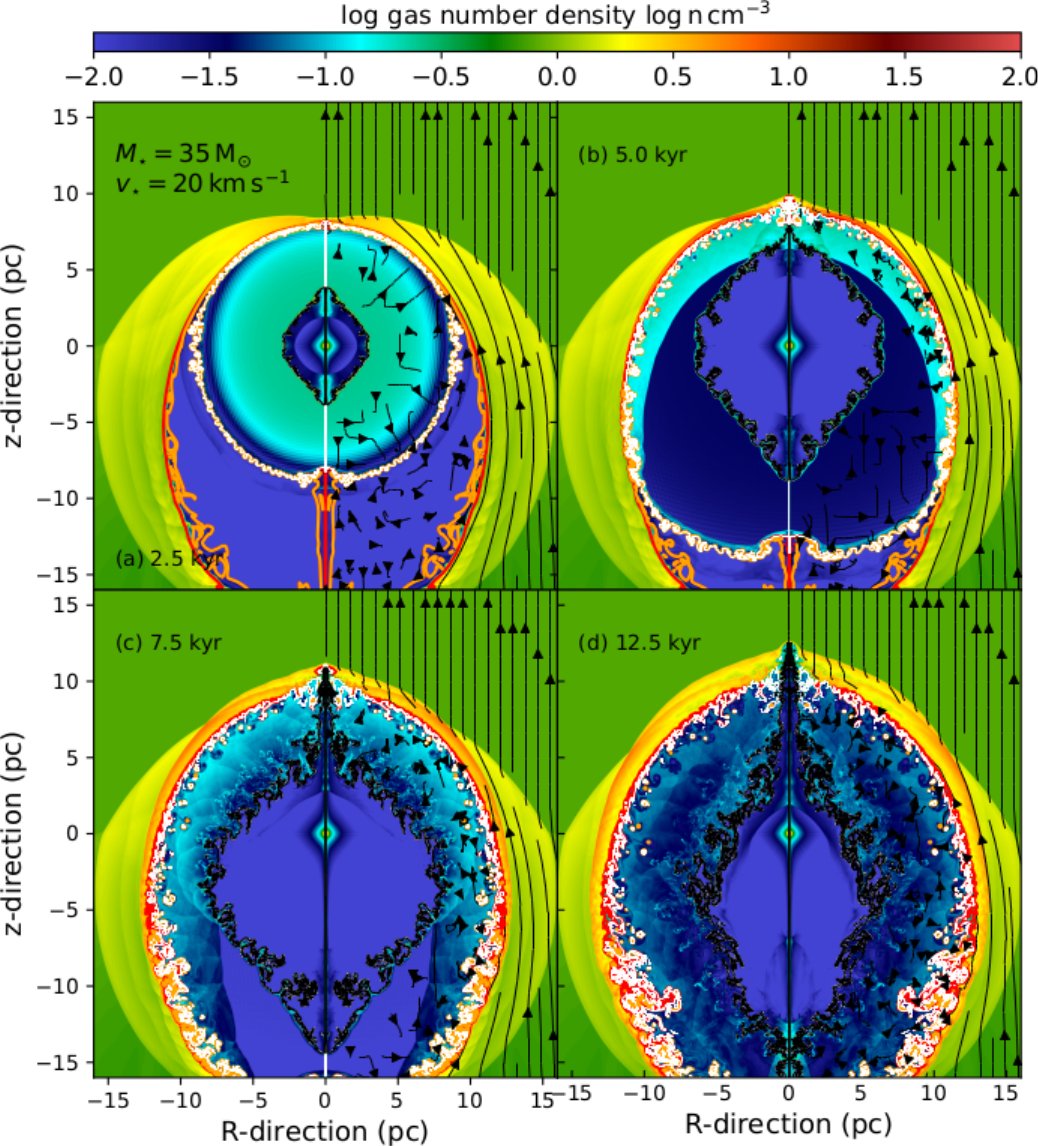}  \\        
        \caption{
        As Fig.~\ref{fig:snr_pwn_2020} for a $35\, \rm M_{\odot}$ progenitor star 
        moving with velocity $v_{\star}=20\, \rm km\, \rm s^{-1}$. 
        }
        \label{fig:snr_pwn_3520}  
\end{figure*}

\begin{figure*}
        \centering
        \includegraphics[width=1.4\columnwidth]{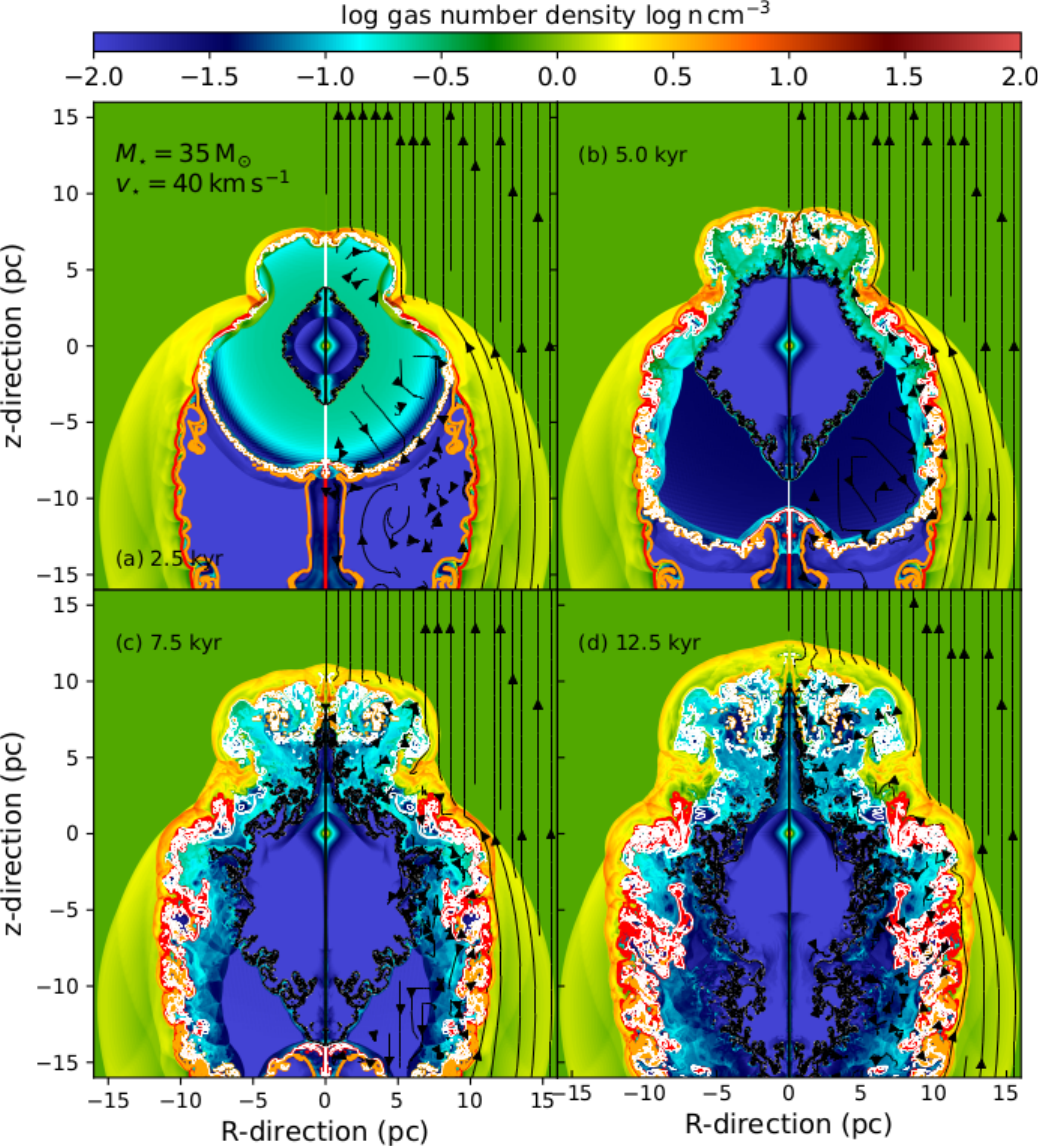}  \\        
        \caption{
        As Fig.~\ref{fig:snr_pwn_2020} for a $35\, \rm M_{\odot}$ progenitor star 
        moving with velocity $v_{\star}=40\, \rm km\, \rm s^{-1}$. 
        }
        \label{fig:snr_pwn_3540}  
\end{figure*}

\begin{figure*}
        \centering
        \includegraphics[width=1.45\columnwidth]{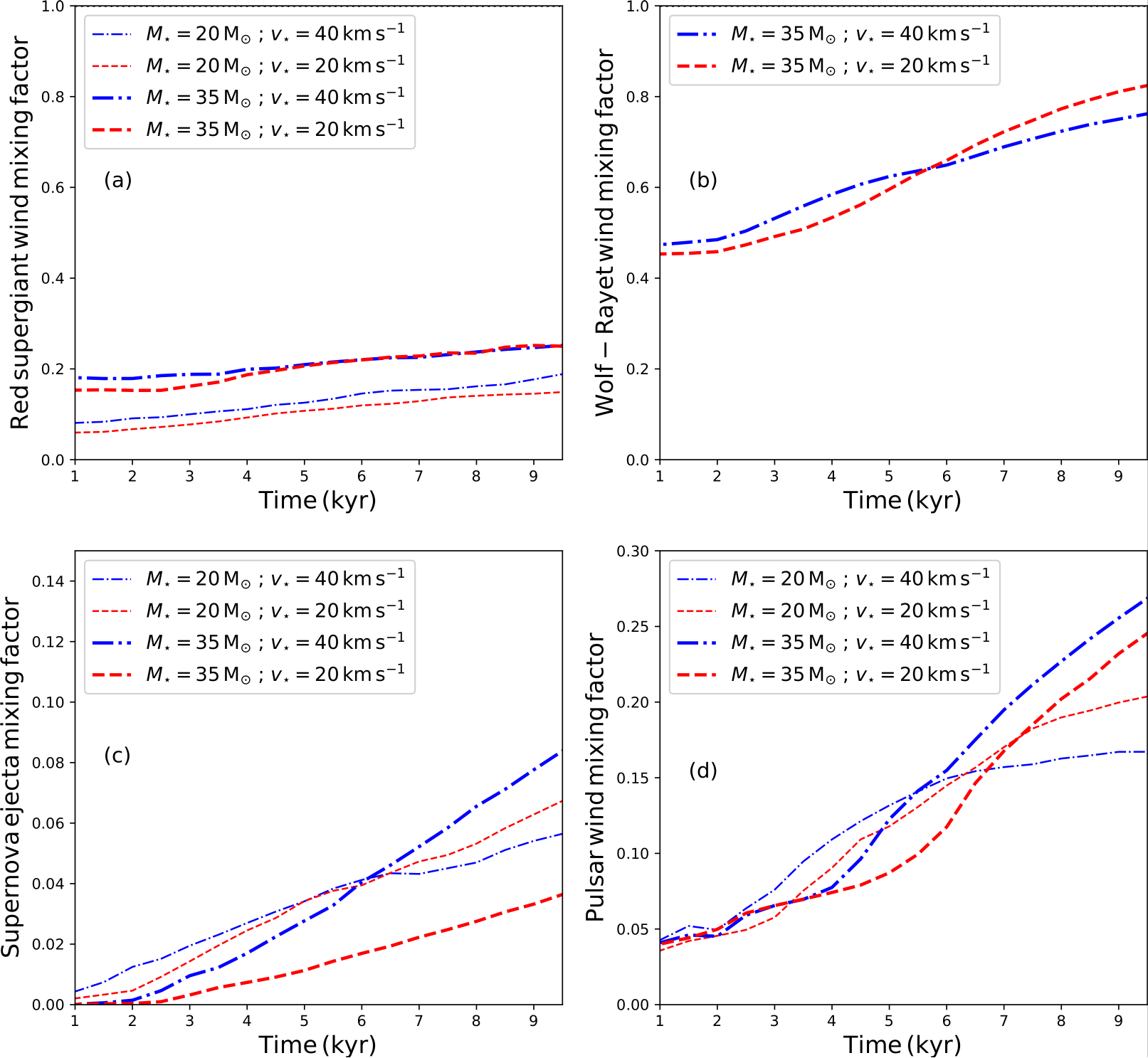}  \\        
        \caption{
        Mixing of the different materials in the plerionic supernova remnant 
        during the first $9\, \rm kyr$ after the supernova explosion. 
        The line colors distinguish models with stellar velocity 
        $v_{\star}=20\, \rm km\, \rm s^{-1}$ (dashed red) and  
        $v_{\star}=40\, \rm km\, \rm s^{-1}$ (dotted dashed blue), 
        and for 
        $M_{\star}=20\, \rm M_\odot$ (thin lines) and  
        $M_{\star}=35\, \rm M_\odot$ (thick lines).   
        The panels display the mixing for the red supergiant (a), 
        Wolf-Rayet (b) wind material, supernova ejecta (c) 
        and pulsar wind (d), respectively. 
        }
        \label{fig:mixing}  
\end{figure*}


\subsection{Initial conditions}
\label{method_ic}

A spherically-symmetric stellar wind is injected onto a sphere of radius 
20 cells centered into the origin of the computational domain, by imposing 
therein mass density and terminal velocity, evaluated from the escape wind 
velocity with the recipe of \citet{eldridge_mnras_367_2006}. 
The stellar wind properties are interpolated from the stellar evolutionary tracks for 
the $20\, \rm M_\odot$ and the $35\, \rm M_\odot$ rotating stars of the 
{\sc geneva} modes \citep{ekstroem_aa_537_2012}, from the zero-age main-sequence 
to the pre-supernova times. 
The main stellar wind properties are reported in Fig. \ref{fig:plot_star_properties}. 
The stellar wind terminal radial velocity is calculated via the relation,  
\begin{equation}    
     v_{\rm w}(t) = \sqrt{ \beta(T)  \frac{  2 G M_{\star}(t) }{ R_{\star}(t)} },
\end{equation}
with $G$, $M_{\star}$ and $R_{\star}$ the gravitational constant, the stellar mass, and  
the radius of the star. The term $\beta(T)$ comes from the recipe of~\citet{eldridge_mnras_367_2006}.  
The circumstellar medium used as initial conditions is calculated assuming that the runaway progenitor 
is a blackbody radiator, i.e. estimating the stellar radius with its photospheric luminosity and 
its effective temperature, respectively, which diminishes the terminal speed during the Wolf-Rayet 
phase of the higher-mass progenitor~\citep{meyer_mnras_521_2023,meyer_aa_687_2024}, making our model 
consistent with the values measures for weak-winded Galactic Wolf-Rayet stars, see the work of 
~\citet{hamman_aa_625_2019}. 
The initial equatorial rotational velocity of 
the progenitors is chosen to $10\%$ of the star's break-up rotation rate. 
The magnetic field of the stars is estimated from observations for OB 
stars~\citep{fossati_aa_574_2015,przybilla_aa_587_2016} and red 
supergiants~\citep{vlemmings_aa_394_2002,kervella_aa_609_2018}, 
surperimposed into the stellar wind ~\citep{rozyczka_apj_469_1996,   
garciasegura_apj_860_2018} and its strength is scaled to that of 
the decrease of the solar wind magnetisation 
as~\citep{scherer_mnras_493_2020,baalmann_aa_634_2020}.

The $10^{51}\, \rm erg$ supernova explosion is modelled using the 
prescription of \citet{chevalier_apj_258_1982,truelove_apjs_120_1999}, 
which is set on a power-law density profile (an inner
dense plateau and a 
decreasing envelope). 
Together with an homologous expansion profile for the velocity, and ejecta masses 
of $7.28\, \rm M_\odot$ and $10.12\, \rm M_\odot$ for the $20\, \rm M_\odot$ 
and $35\, \rm M_\odot$ progenitors, we implemented the onset of 
the blaswtave as described in~\citet{whalen_apj_682_2008,vanveelen_aa_50_2009}.

The profiles for the blastwave reads 
\begin{equation}
\rho(r) = \begin{cases}
        \rho_{\rm core}(r) & \text{if $r \leq r_{\rm core}$},               \\
        \rho_{\rm max}(r)  & \text{if $r_{\rm core} < r < r_{\rm max}$},    \\
        \end{cases}
	\label{cases}
\end{equation}
where, 
\begin{equation}
   \rho_{\rm core}(r) =  \frac{1}{ 4 \pi n } \frac{ (10 E_{\rm ej}^{n-5})^{-3/2}
 }{  (3 M_{\rm ej}^{n-3})^{-5/2}  } \frac{ 1}{t_{\rm max}^{3} },
   \label{sn:density_1}
\end{equation}
and, 
\begin{equation}
   \rho_{\rm max}(r) =  \frac{1}{ 4 \pi n } \frac{ \left(10 E_{\rm
ej}^{n-5}\right)^{(n-3)/2}  }{  \left(3 M_{\rm ej}^{n-3}\right)^{(n-5)/2}  } \frac{ 1}{t_{\rm max}^{3} } 
\bigg(\frac{r}{t_{\rm max}}\bigg)^{-n},
   \label{sn:density_2}
\end{equation}
and with the ejecta mass $M_{\rm ej}$, the ejecta energy $E_{\rm ej}$ 
and $n=11$~\citep{truelove_apjs_120_1999,bandiera_mnras_508_2021}. 
Additionally, 
\begin{equation}
    t_{\rm max} = \frac{r_{\rm max}}{v_{\rm max}},
\end{equation}
is calculated following~\citet{whalen_apj_682_2008}. 
The velocity is set as $ v(r) = r/t$ with 
\begin{equation}
   v_{\mathrm{core}} = \bigg(  \frac{ 10(n-5)E_{\mathrm{ej}} }{ 3(n-3)M_{\mathrm{ej}} } \bigg)^{1/2},
   \label{sn:vcore}
\end{equation}
and 
\begin{equation}
v_{\mathrm{max}}= \frac{r_{\mathrm{max}}}{t_{\mathrm{max}}}=3\times 10^{4}\, \mathrm{km\, s^{-1}},
\end{equation}
\textcolor{black}{
with $t_{\rm max}=5.11\, \times 10^{-8}\, \rm Myr$ and $r=r_{\mathrm{max}}=0.00153\, \rm pc$.   
The radii marking the end of the blastwave plateau is 
$r=r_{\mathrm{core}}=0.000212\, \rm pc$ for 
$v_{\mathrm{core}}=4155.2\, \mathrm{km\, s^{-1}}$ in the case of the $20\, \rm M_\odot$
and $r=r_{\mathrm{core}}=0.000180\, \rm pc$ for 
$v_{\mathrm{core}}=3524.4\, \mathrm{km\, s^{-1}}$ in the case of the $35\, \rm M_\odot$, 
respectively. 
}

\textcolor{black}{
The pulsar wind is set on the prescription of~\citet{komissarov_mnras_349_2004}. 
We assume the pulsar to be of initial power,
\begin{equation}
\dot{E}(t) = \dot{E}_{\mathrm{o}} \left(1 + \frac{t}{\tau_{\mathrm{o}}} \right)^{\alpha},
\end{equation}
with,  
\begin{equation}
\alpha = \frac{  n+1 }{  n-1 },
\end{equation}
where n is the braking index of the pulsar
and where the initial spin-down is given by, 
\begin{equation}
\tau_{\mathrm{o}} = \frac{P_{\mathrm{o}}}{(n-1)\dot{P}_{\mathrm{o}}},
\end{equation}
where following \citet{2017hsn_book_2159S}. 
We set $\dot{E}_{\mathrm{o}} = 10^{38}\,\mathrm{erg}\,\mathrm{s}^{-1}$, 
with a time-dependence governed with $n=3$, a period 
$P_{\mathrm{o}} = 0.3\,\mathrm{s}$ and a period derivative 
$\dot{P}_{\mathrm{o}} = 10^{-17}\,\mathrm{s}\,\mathrm{s}^{-1}$. 
The relativistic pulsar wind is of velocity $0.01\, \rm c$ 
\citep{2017hsn_book_2159S} and its corresponding mass density 
is estimated with the laws presented in \citet{swaluw_aa_404_2003}. 
The magnetic field associated to the pulsar wind is injected onto a 
sphere of radius $20$ cells as, 
\begin{equation}
B_{\mathrm{psr}}(r,t) = 
\sqrt{ 4\pi \frac{ \dot{E}(t) }{ v_{ \mathrm{psr} }  }}  
\frac{\sqrt{  \sigma }}{r}  
\sin(\theta)\left(1 - \frac{2\theta}{\pi}\right),
\end{equation}
where $\sigma=10^{-3}$ \citep{2017hsn_book_2159S} and 
$v_\mathrm{psr}=0.01c$ the pulsar wind speed where $c$ 
is the speed of light. 
}

The calculations are performed in the frame of the moving progenitor, 
assuming two distinct bulk velocities $v_{\star}=20\, \rm km\, \rm s^{-1}$
and $v_{\star}=40\, \rm km\, \rm s^{-1}$. 
We summarise the models in this study, together with their corresponding 
evolutionary sequence, in Table \ref{tab:table1}.

\subsection{Numerical methods}
\label{method_num}

We carry out our simulations within the frame of the magneto-hydrodynamics with the 
{\sc pluto} code, presented in \citet{mignone_apj_170_2007,migmone_apjs_198_2012}, 
by solving the following set of equations: 
\begin{equation}
    \frac{\partial \rho}{\partial t} + 
    \vec{\nabla} \cdot (\rho\vec{v}) = 0,
    \label{eq:mhdeq_1}
\end{equation}
\begin{equation}
    \frac{\partial \vec{m}}{\partial t} +
    \vec{\nabla} \cdot \left( \vec{m} \otimes \vec{v}  
    - \vec{B} \otimes \vec{B} + \hat{\vec{I}}p_{\rm t} \right)
    = \vec{0},
    \label{eq:mhdeq_2}
\end{equation}
\begin{equation}
    \frac{\partial E}{\partial t} + 
    \vec{\nabla} \cdot \left( (E+p_{\rm t})\vec{v}-\vec{B}(\vec{v}\cdot\vec{B}) \right)
    = \Phi(T,\rho),
    \label{eq:mhdeq_3}
\end{equation}
\begin{equation}
    \frac{\partial \vec{B}}{\partial t} + 
    \vec{\nabla} \cdot \left( \vec{v} \otimes \vec{B} - \vec{B} \otimes \vec{v} \right)
    = \vec{0},
    \label{eq:mhdeq_4}
\end{equation}
with $\rho$ is the mass density, $\vec{v}$ the velocity vector,
$\vec{m}=\rho\vec{v}$ the momentum vector, $\hat{\vec{I}}$ 
the identity vector, $\vec{B}$ the magnetic field vector, 
$p_{\rm t}=p+\vec{B}^{2}/8\pi$ the total pressure, and, 
\begin{equation}
    E = \frac{p}{(\gamma - 1)} + \frac{\vec{m} \cdot \vec{m}}{2\rho} + \frac{\vec{B} \cdot \vec{B}}{2},
    \label{eq:energy}
\end{equation}
the total energy, where $\gamma=5/3$ the adiabatic index. 
The system is evolved under the assumption of an ideal equation of state and 
the governing equations are closed with,
\begin{equation}
	  c_{\rm s} = \sqrt{ \frac{\gamma p}{\rho} },
\end{equation}
where $c_{\rm s}$ is the sound speed of the plasma. The term 
\begin{equation}  
    \Phi(T,\rho) = n_{\mathrm{H}}\Gamma(T) - n^{2}_{\mathrm{H}}\Lambda(T),
    \label{eq:dissipation}
\end{equation}
represents the optically-thin radiative processes such as cooling and heating, 
where,  
\begin{equation}
    T = \mu \frac{m_{\mathrm{H}}}{k_{\rm{B}}} \frac{p}{\rho},
    \label{eq:temperature}
\end{equation}
is the gas temperature, see \citet{meyer_2014bb}. This terms are not included   
starting from the launching of the pulsar wind. After that moment, the system is 
evolved adiabatically.  
%

The bow shock constituting the circumstellar medium of the runaway core-collapse 
progenitor is modelled using a 2.5-dimensional cylindrical coordinate system $(R,z)$. 
The size of the utilised computational domain is $[0;100]\times[-50;50]\, \rm pc^{2}$ 
and it is mapped with a $2000 \times 2000$ grid zones, i.e. with a uniform spatial 
resolution of $\Delta=5.0 \times 10^{-2} \, \rm pc$. 
\textcolor{black}{
The results for the circumstellar medium are displayed in Fig. \ref{fig:csm_at_psn}. 
}
The initial interaction between the supernova blastwave and the freely-expanding 
last stellar wind of the progenitor is calculated with a 1-dimensional spherical 
computational domain, mapped with a $[0;2.5]\, \rm pc^{2}$ that is uniformly 
discretised with $125000$ grid zones, which is equivalent to a spatial resolution 
of $\Delta=2.0 \times 10^{-5} \, \rm pc$. 
The simulations for the supernova remnants are performed with a 2.5-dimensional 
cylindrical coordinate system $(R,z)$ (i.e., a 2 dimensional cylindrical grid
for the scalar quantities plus a toroidal component for the vectors) onto which 
a $[0;25]\times[-25;25]\, \rm pc^{2}$ 
computational domain is mapped with a $4000 \times 8000$ grid zones, permitting 
to reach everywhere spatial resolution of $\Delta=6.25 \times 10^{-3} \, \rm pc$.

We make use of a finite-volume, dimensionally unsplit Godunov-type solver that 
is made of the HLL Rieman solver \citep{hll_ref} together with the third-order 
Runge-Kutta integrator for the time-marching algorithm. The spatial reconstruction 
of the variables between the grid zones are performed with the piecewise parabolic 
method of interpolation for the models for the circumstellar medium, and are 
performed with a linear spatial reconstruction and the minmod slope limiter 
for the calculations for the supernova remnants, making the later scheme total 
variation diminishing.  
The timesteps are controlled by the Courant-Friedrich-Levy number which 
we initially set to 0.3. The scheme continuously ensures the divergence-free 
character of the magnetic field vector everywhere in the computational 
domain \citep{Powell1997}. 
%
%
We refer interested readers 
on further details about the the numerical method 
to the previous studies \citet{meyer_527_mnras_2024,meyer_aa_687_2024}.


\section{Results}
\label{results}

%

\subsection{Distribution of material in the pulsar wind nebulae}
\label{resultats_distribution}

Fig.~\ref{fig:snr_pwn_2020} plots the number density fields in our 
magneto-hydrodynamical simulation of the supernova remnant of a 
runaway $20\, \rm M_{\odot}$ star, initially rotating with 
$\Omega_{\star}/\Omega_{\rm K}=0.1$ and moving with velocity 
$v_{\star}=20\, \rm km\, \rm s^{-1}$. Its evolution is shown 
at times $2.5$ (a), $5.0$ (b), $7.5$ (c) and $12.5\, \rm kyr$ (d), 
respectively. The contours mark the region with a $50\%$ contribution of 
supernova ejecta (white), red supergiant wind (red) and with a $10\%$ 
contribution of main-sequence material (cyan), respectively. 
The black arrows in the right-hand parts of the figures are ISM 
magnetic field lines. Fig.~\ref{fig:snr_pwn_2040} is similar as 
Fig.~\ref{fig:snr_pwn_2020} for a progenitor velocity of 
$v_{\star}=40\, \rm km\, \rm s^{-1}$ and Figs.~\ref{fig:snr_pwn_3520}, 
~\ref{fig:snr_pwn_3540} are for a $35\, \rm M_{\odot}$ star with 
the Wolf-Rayet wind highlighted in orange.

The supernova remnant in the model with $20\, \rm M_{\odot}$ star and 
$v_{\star}=20\, \rm km\, \rm s^{-1}$ (Fig.~\ref{fig:snr_pwn_2020}a) 
has the overall shape of a Napoleon's hat, when the blastwave remains 
partly trapped into the red supergiant bow shock, which distributes as 
an arc/shell of material interacting with the main-sequence material (red). 
The low-density cavity behind the star is filled with main-sequence wind 
(cyan). The supernova material expands spherically first, then interacts 
asymetrically with the circumstellar medium (white), as described 
in \citet{meyer_mnras_450_2015}. The pulsar wind nebula grows inside of 
the ejecta as reported in \citet{swaluw_aa_404_2003,komissarov_mnras_349_2004}. 
When the remnant growth, the supernova ejecta are filling a ring which outer 
border is the shocked dense circumstellar medium, and the inner border is 
the termination shock of the pulsar wind nebula which entered reverberation. 
Because of the density and velocity differences between the species, 
Rayleight-Taylor instabilities develop in the ejecta ring 
(Fig.~\ref{fig:snr_pwn_2020}b). As the shock wave is channeled into 
the trail of the bow shock, the pulsar wind nebula adopts an oblong 
shape in which the leptonic wind develops strong instabilities with 
the other materials downstream the location of the explosion. 
The supernova remnant in the model with $20\, \rm M_{\odot}$ star and 
$v_{\star}=40\, \rm km\, \rm s^{-1}$ (Fig.~\ref{fig:snr_pwn_2040}) 
is the model that corresponds to a Cygnus-Loop-like supernova 
remnant \citep{meyer_527_mnras_2024} and which evolution is 
presented in \citet{meyer_mnras_515_2022}. 
The ring of expanding ejecta has the drop-like morphology outflowing 
from the bow shock, quickly affected by strong instabilities at the 
ejecta/supergiant discontinuity and inducing important deviation of the 
pulsar wind nebulae from the solution of \citet{komissarov_mnras_349_2004}.

Since the cavity carved by the Wolf-Rayet wind of the $35\, \rm M_{\odot}$ 
star is larger than that of the red supergiant wind alone by the 
$20\, \rm M_{\odot}$ star, then the ring of supernova ejecta expands 
freely in the stellar wind.
This happens until reaching distances of $5\, \rm pc$ 
when it begins to interact with the shell of mixed Wolf-Rayet and red 
supergiant stellar wind. The pulsar wind develops as a diamond shape 
inside the ejecta (Fig.~\ref{fig:snr_pwn_3520}a). 
At the time of the reflection of the supernova blastwave with the 
circumstellar medium ($5.0\, \rm kyr$), the pulsar wind is still 
unperturbed in its expansion by the external medium 
(Fig.~\ref{fig:snr_pwn_3520}b) and then it begins the reverberation 
phase, that is induced by the reflected blastwave in the northern 
region of the supernova remnant (Fig.~\ref{fig:snr_pwn_3520}c), 
accelerating the mixing of pulsar wind material and supernova ejecta 
in the remnant, while that of the supernova ejecta with the stellar 
wind diminishes. The eddies of the instabilities at the blastwave-wind 
interface are much smaller than that in the model with $20\, \rm M_{\odot}$. 
When the $35\, \rm M_{\odot}$ progenitor moves with velocity 
$v_{\star}=40\, \rm km\, \rm s^{-1}$, the circumstellar medium 
is smaller and more much complex, which profoundly modifies the 
reflection of the blastwave (Fig.~\ref{fig:snr_pwn_3540}a). 
The breaking of the diamond shape of the pulsar wind nebula due to the 
reflection of the blastwave is more pronounced, both along and opposite 
of the direction of motion of the progenitor (Fig.~\ref{fig:snr_pwn_3540}b,c). 
The reflection is more efficient and travels back closer to the center 
of the explosion than when $v_{\star}=20\, \rm km\, \rm s^{-1}$, creating 
a thin region of mixed stellar winds and supernova ejecta surrounding 
the irregular pulsar wind nebula.

\subsection{Mixing of materials}
\label{resultats_mixing}

Fig.~\ref{fig:mixing} displays the evolution of the mixing efficiency of the 
post-main-sequence stellar winds, i.e. red supergiant (a) and Wolf-Rayet (b), 
the supernova ejecta (c) and the pulsar wind (d). The quantities are plotted 
for the $20\, \rm M_{\odot}$ (thin lines) and the  $35\, \rm M_{\odot}$ progenitor 
stars (thick lines), moving with velocities $v_{\star}=20\, \rm km\, \rm s^{-1}$ 
(dashed lines) and $v_{\star}=40\, \rm km\, \rm s^{-1}$ (dashed dotted lines). 
The definition of the mixing efficiency $f_i$ of a given specie that we adopt 
follows the prescription of \citet{orlando_aa_444_2005}. They read, 
\begin{equation}
        f_{\rm MS} =   \frac{ \mu m_{\rm H} \iint_{ Q_{\rm MS}\le 0.5 }  
        n Q_{\rm MS} \mathrm{dV} }{ \mu m_{\rm H} \iint_{\mathrm{SNR}}   n Q_{\rm MS} \mathrm{dV} }
          =   \frac{ \iint_{ Q_{\rm MS}\le 0.5 }  
        n Q_{\rm MS} \mathrm{dV} }{ \iint_{\mathrm{SNR}}   n Q_{\rm MS} \mathrm{dV} },
        \label{eq:f_MS}  
\end{equation}
\begin{equation}
        f_{\rm RSG}  =   \frac{ \iint_{ Q_{\rm RSG}\le 0.5 }  
        n Q_{\rm RSG} \mathrm{dV} }{ \iint_{\mathrm{SNR}}   n Q_{\rm RSG} \mathrm{dV} },
        \label{eq:f_RSG}  
\end{equation}
\begin{equation}
        f_{\rm WR}  =   \frac{ \iint_{ Q_{\rm WR}\le 0.5 }  
        n Q_{\rm WR} \mathrm{dV} }{ \iint_{\mathrm{SNR}}   n Q_{\rm WR} \mathrm{dV} },
        \label{eq:f_WR}  
\end{equation}
\begin{equation}
        f_{\rm EJ} 
          =   \frac{ \iint_{ Q_{\rm EJ}\le 0.5 }  
        n Q_{\rm EJ} \mathrm{dV} }{ \iint_{\mathrm{SNR}}   n Q_{\rm EJ} \mathrm{dV} },
        \label{eq:f}  
\end{equation}
\begin{equation}
        f_{\rm PSR} 
          =   \frac{ \iint_{ Q_{\rm PSR}\le 0.5 }  
        n Q_{\rm PSR} \mathrm{dV} }{ \iint_{\mathrm{SNR}}   n Q_{\rm PSR} \mathrm{dV} },
        \label{eq:f}  
\end{equation}
respectively, 
where $\rm dV$ is the volume element of the $2.5$-dimensional supernova remnant model. 
The integrals are performed onto the supernova remnants volume that is selected from 
the rest of the computational domain using a criterion based on the number density 
of the unperturbed ambient medium. The denominator of each mixing factors represents 
the total mass of a given specie inside of the supernova remnant, and the numerator 
stands for the mass of a given specie in the cells that are filled with less than 
$50\%$ in number density of that particular kind of material. 
Hence $f_\mathrm{i}=0$ means that there is no mixing happening in the supernova 
remnant, i.e. that it is, in the average, pure with respect to that particular material 
$i \in \{ \mathrm{MS}, \mathrm{RSG}, \mathrm{WR}, \mathrm{EJ}, \mathrm{PSR} \}$, 
in the sense that most of its mass is made of that particular specie. 
When $f_\mathrm{i}=1$, that material is diluted and it contributes by $\le 50$ 
per cent in number density to the local chemical composition of the gas, see 
also \citet{meyer_mnras_521_2023}.

The total amount of red supergiant wind mixes by $\le 20\%$, mostly due to the 
fact that large quantities of this material constitute the dense circumstellar 
medium and has filled the low-density trail of the runaway progenitor's bow shock, 
which are not yet impacted by the expansion of the supernova blastwave, regardless 
of the progenitor's mass and/or bulk velocity. 
The distribution of Wolf-Rayet material is also strongly affected by the morphology 
of the circumstellar medium, which is rather circular if the $35\, \rm M_\odot$ 
progenitor star moves with velocity $20\, \rm km\, \rm s^{-1}$ as a result of 
the Wolf-Rayet shell forming prior to the explosion \citep{brighenti_mnras_273_1995} 
while this region is more oblong 
if the star move faster. The Wolf-Rayet wind 
mixes efficiently by $50\%$ to $80\%$ within the $10\, \rm  kyr$ after the explosion.

In this study, we will not estimate the mixing of main-sequence material into 
the supernova remnant, since this specie is not present anymore into the stellar 
wind bow shock of the progenitor for the age that we consider ($\simeq 10\, \rm 
kyr$ after the supernova). 
The mixing efficiency relative to the red supergiant material evolves from 
$10\%$ to $15\%$ in the time windows that we consider, showing a $50\%$ difference 
with that in the model with $20\, \rm M_{\odot}$ and $35\, \rm M_{\odot}$
(Fig.~\ref{fig:mixing}a).

The effects of the progenitor 
bulk motion is much milder and even totally vanishes in the $35\, \rm M_{\odot}$ 
model at times $> 5\, \rm kyr$. 
The mixing efficiency of Wolf-Rayet material is only calculated for half of our 
simulations, since such specie is evidently not present in the evolution of the 
initial $20\, \rm M_{\odot}$ star (Fig.~\ref{fig:mixing}b).  
The Wolf-Rayet material is the first specie to be immediately shocked by the 
blastwave, hence, its mixing efficiency increases quicker and already reaches 
$50\%$ at time $1\, \rm kyr$ after the supernova, to have values of early
$80\%$ at time $10\, \rm kyr$, respectively. 
The Wolf-Rayet material mixes better in the model with velocity 
$40\, \rm km\, \rm s^{-1}$ until the remnant ages $5.5\, \rm kyr$, 
then the mixing in that with velocity $20\, \rm km\, \rm s^{-1}$ 
becomes more important then.


\begin{figure*}
        \centering
        \includegraphics[width=1.80\columnwidth]{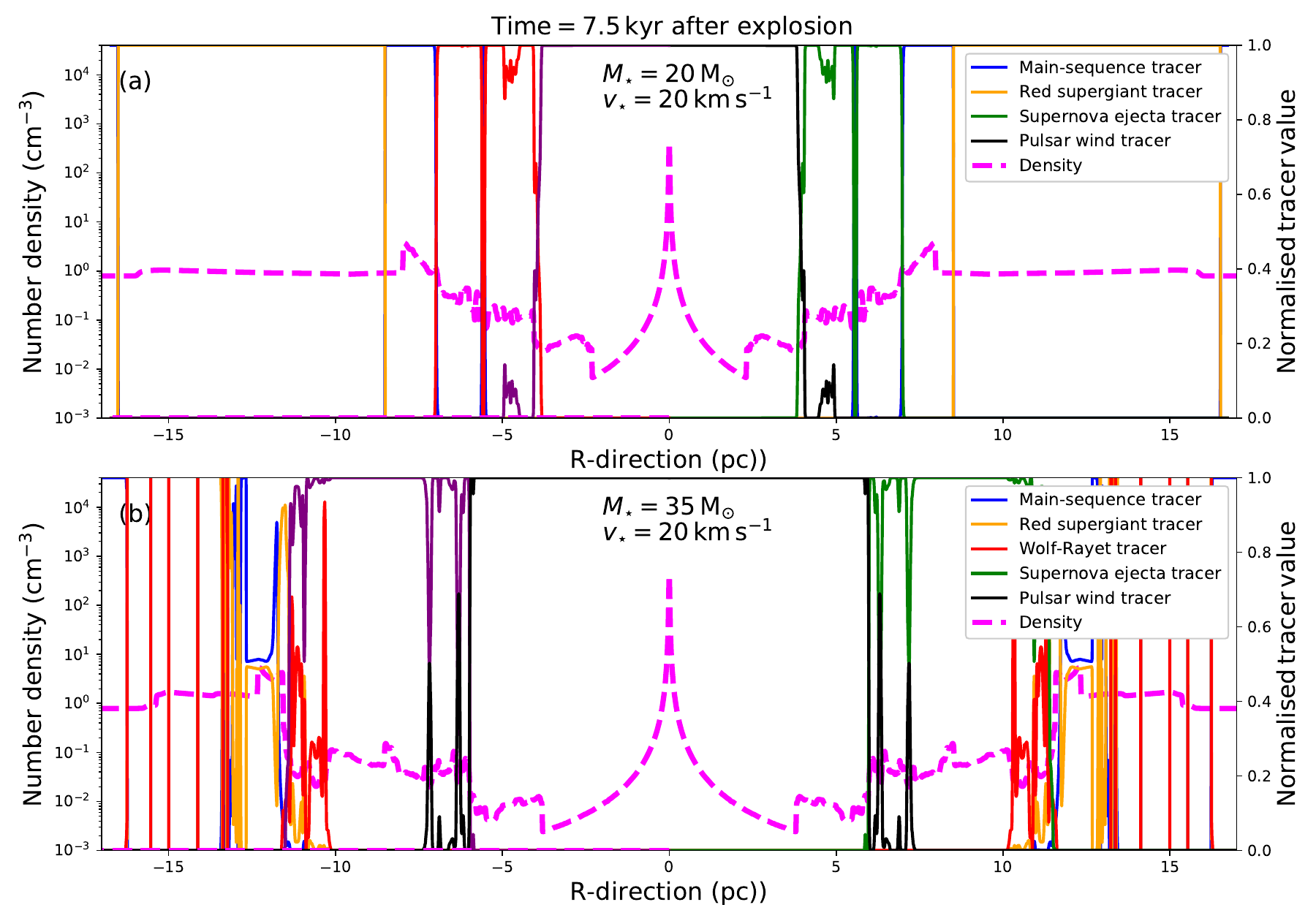}  \\     
        \includegraphics[width=1.80\columnwidth]{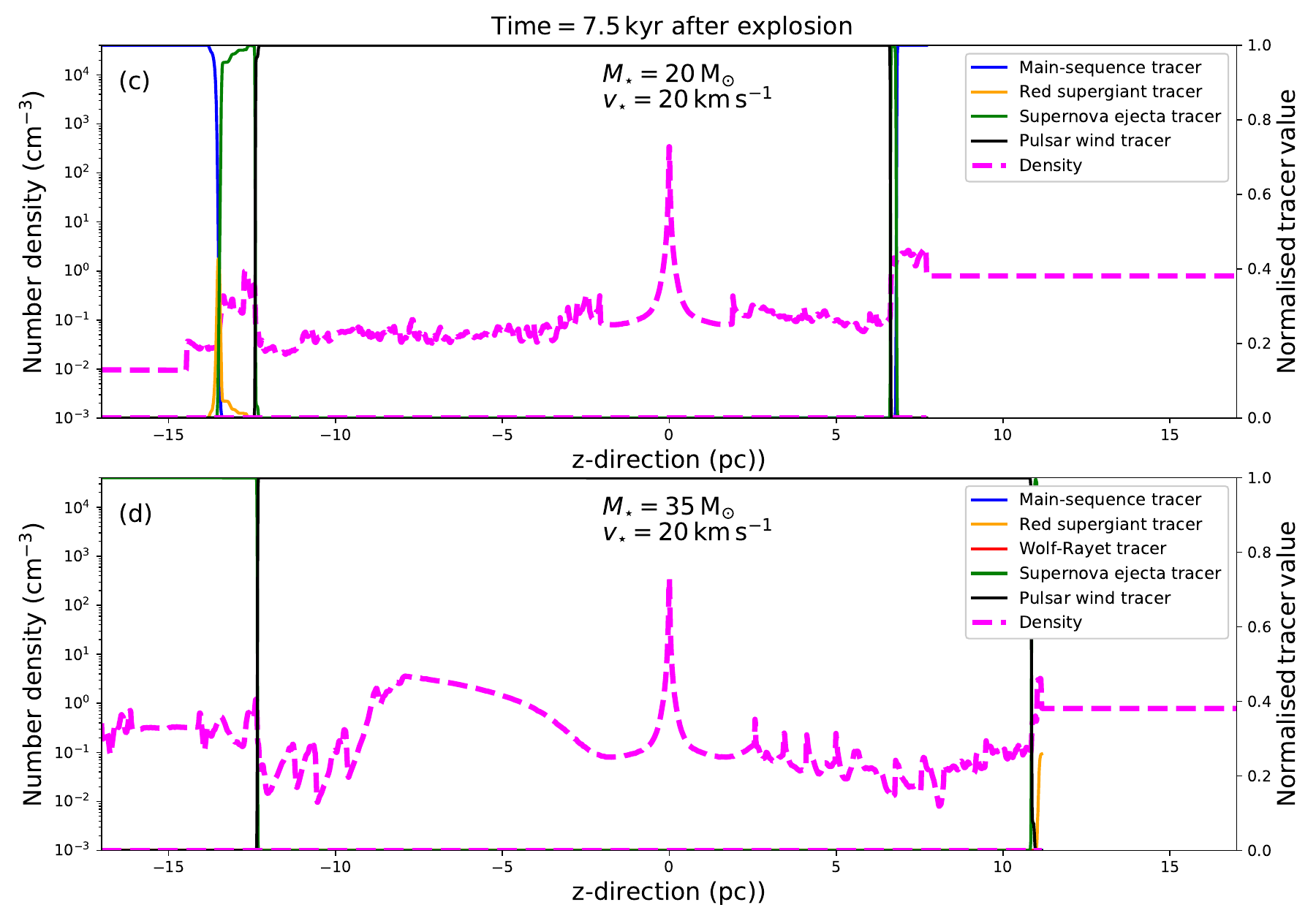}  \\          
        \caption{
        Cuts taken through the supernova remnants generated by the supernova 
        progenitor moving with $v_{\star}=20\, \rm km\, \rm s^{-1}$ prior to 
        the explosion. The slices are along the R-direction direction (top figure) 
        an along the z-direction (bottom figure). In each figure, the panel (a) 
        concerns the $20\, \rm M_{\odot}$ progenitor star and the panel (b)  
        concerns the $35\, \rm M_{\odot}$ progenitor star, respectively. 
        The panels represent the number density (dashed purple line, in $cm^{-3}$), 
        the main-sequence (cyan), red supergiant (orange), Wolf-Rayet (red), 
        supernova ejecta (green) and pulsar wind (black) material tracers. 
        %
        }
        \label{fig:cuts_tracers_20kms}  
\end{figure*}

\begin{figure*}
        \centering
        \includegraphics[width=1.80\columnwidth]{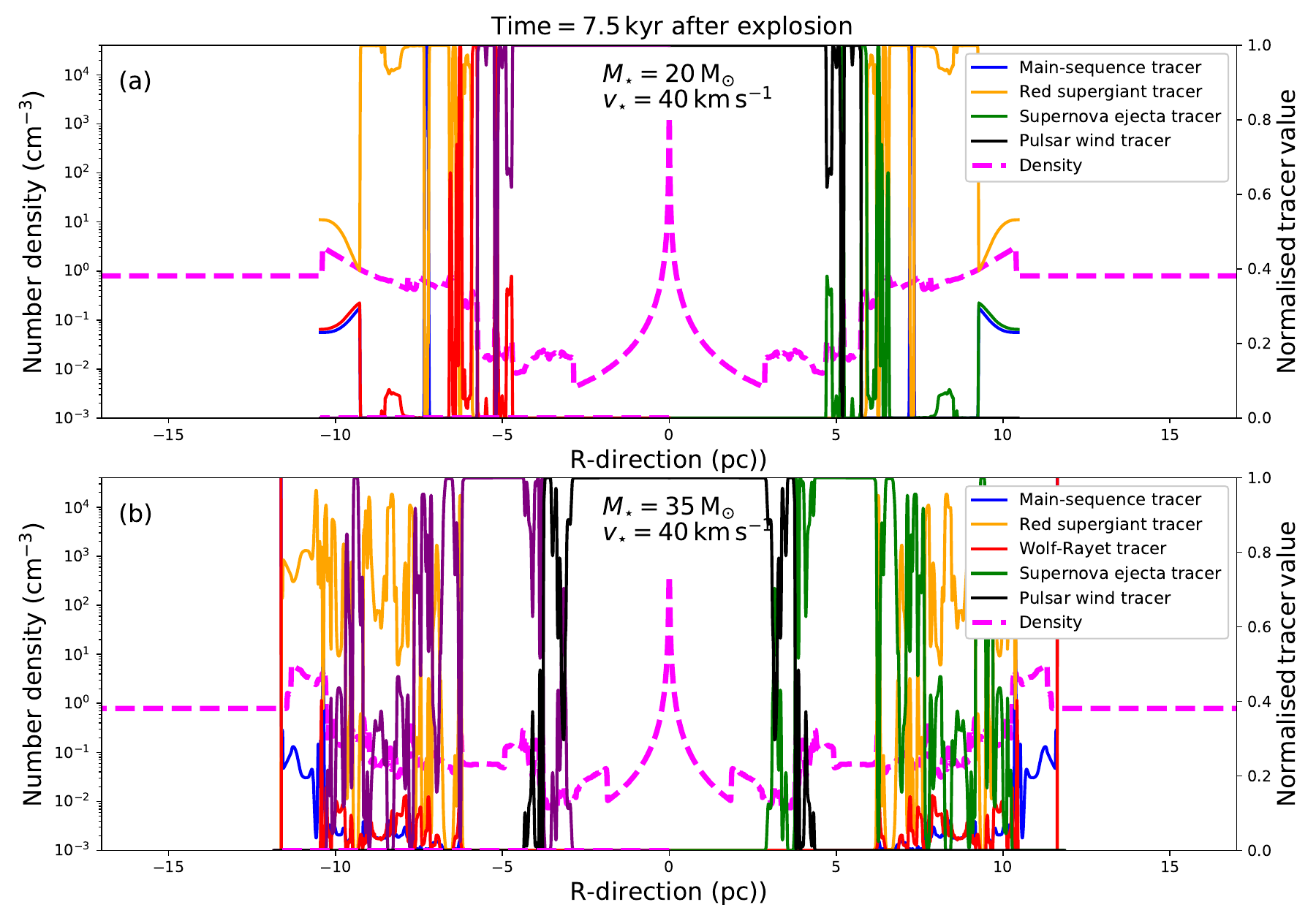}  \\     
        \includegraphics[width=1.80\columnwidth]{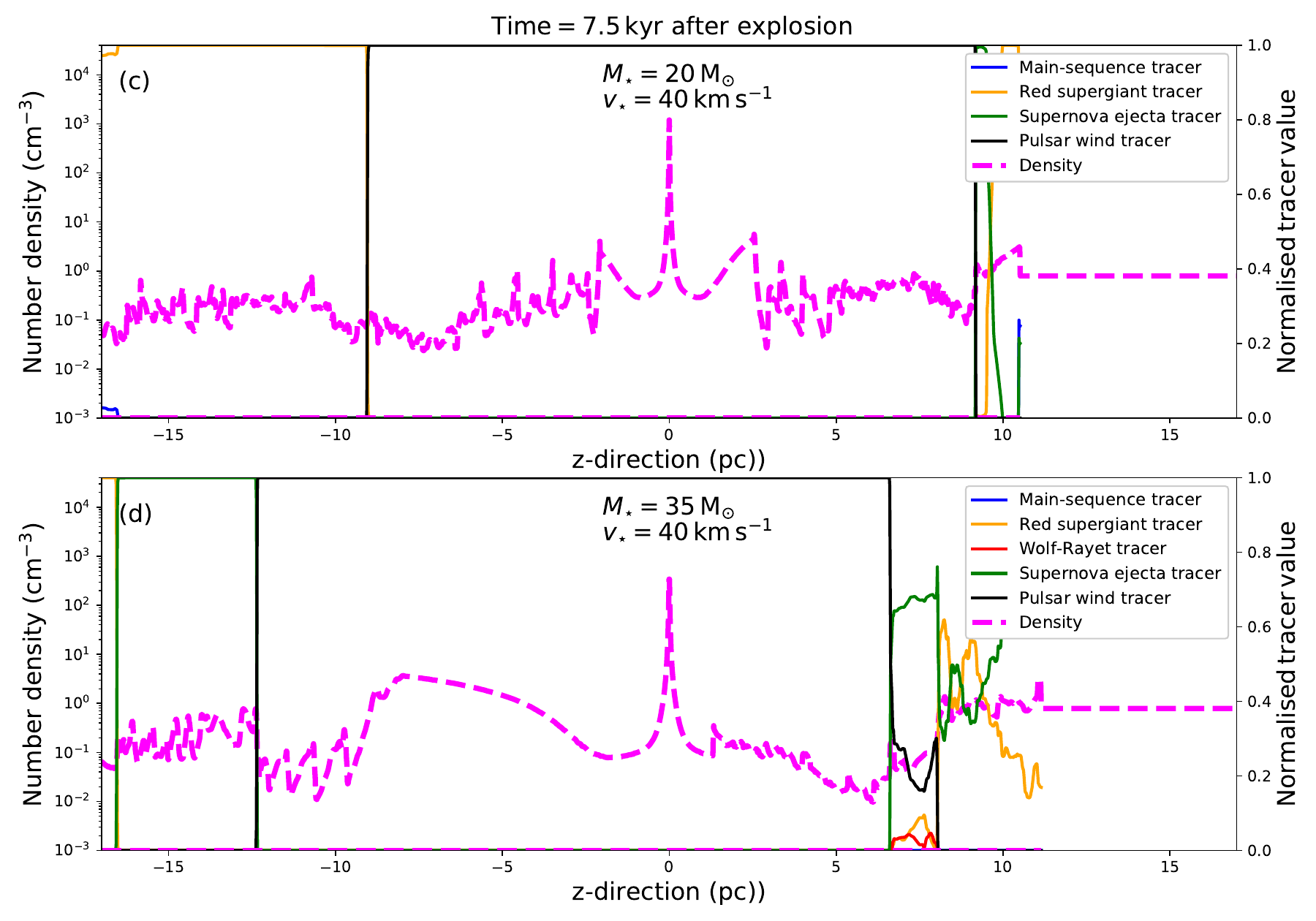}  \\           
        \caption{
        As Fig. \ref{fig:cuts_tracers_20kms} for progenitor stars moving with 
        velocity $v_{\star}=40\, \rm km\, \rm s^{-1}$ prior to the explosion. 
        }
        \label{fig:cuts_tracers_40kms}  
\end{figure*}

\begin{figure*}
        \centering
        \includegraphics[width=1.5\columnwidth]{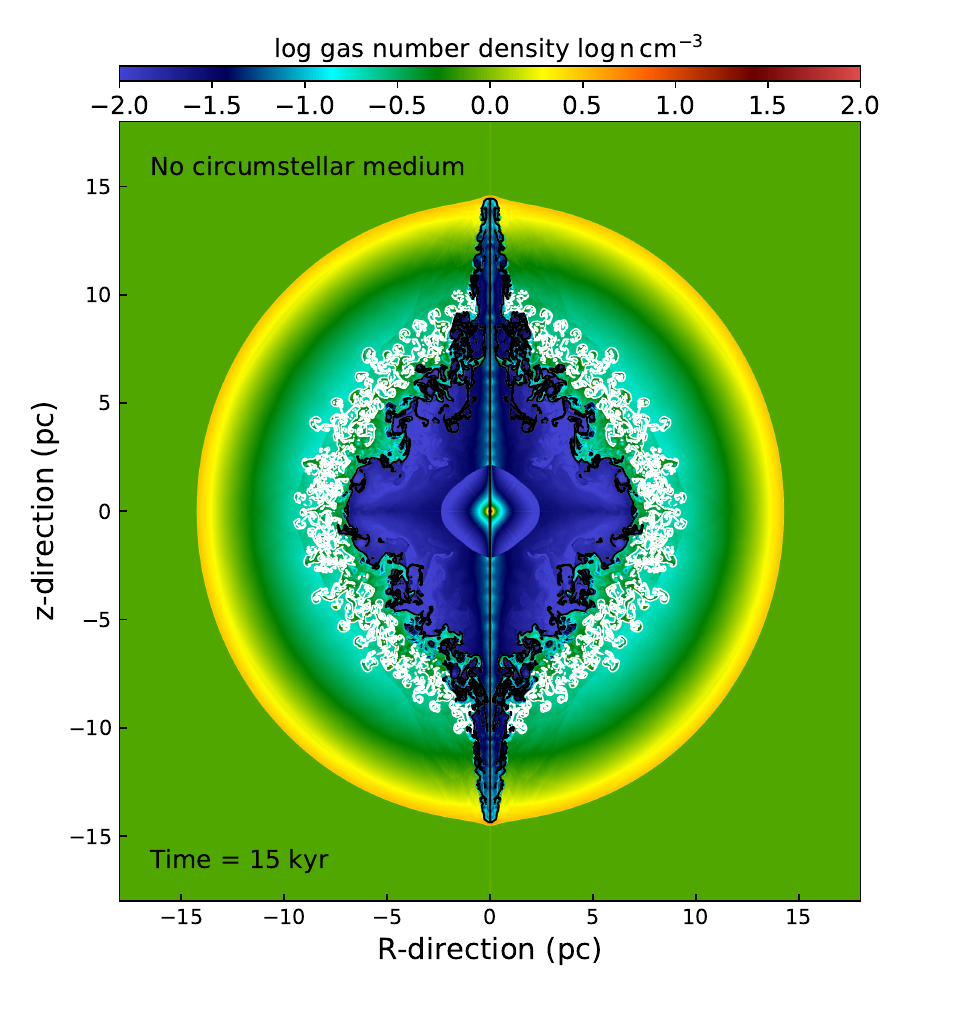}  \\        
        \caption{
        Number density field in a magneto-hydrodynamical simulation of 
        the supernova remnant of a $20\, \rm M_{\odot}$ star rotating 
        with $\Omega_{\star}/\Omega_{\rm K}=0.1$ and moving with velocity 
        $v_{\star}=20\, \rm km\, \rm s^{-1}$. 
        The evolution of the plerionic supernova remnant is shown at 
        $15\, \rm kyr$ after the explosion. 
        The various contours highlight the region with a $50\%$ 
        contribution of pulsar wind (black) and supernova ejecta 
        (white).  
        }
        \label{fig:mhd_model_no_csm}  
\end{figure*}

\begin{figure*}
        \centering
        \includegraphics[width=2.0\columnwidth]{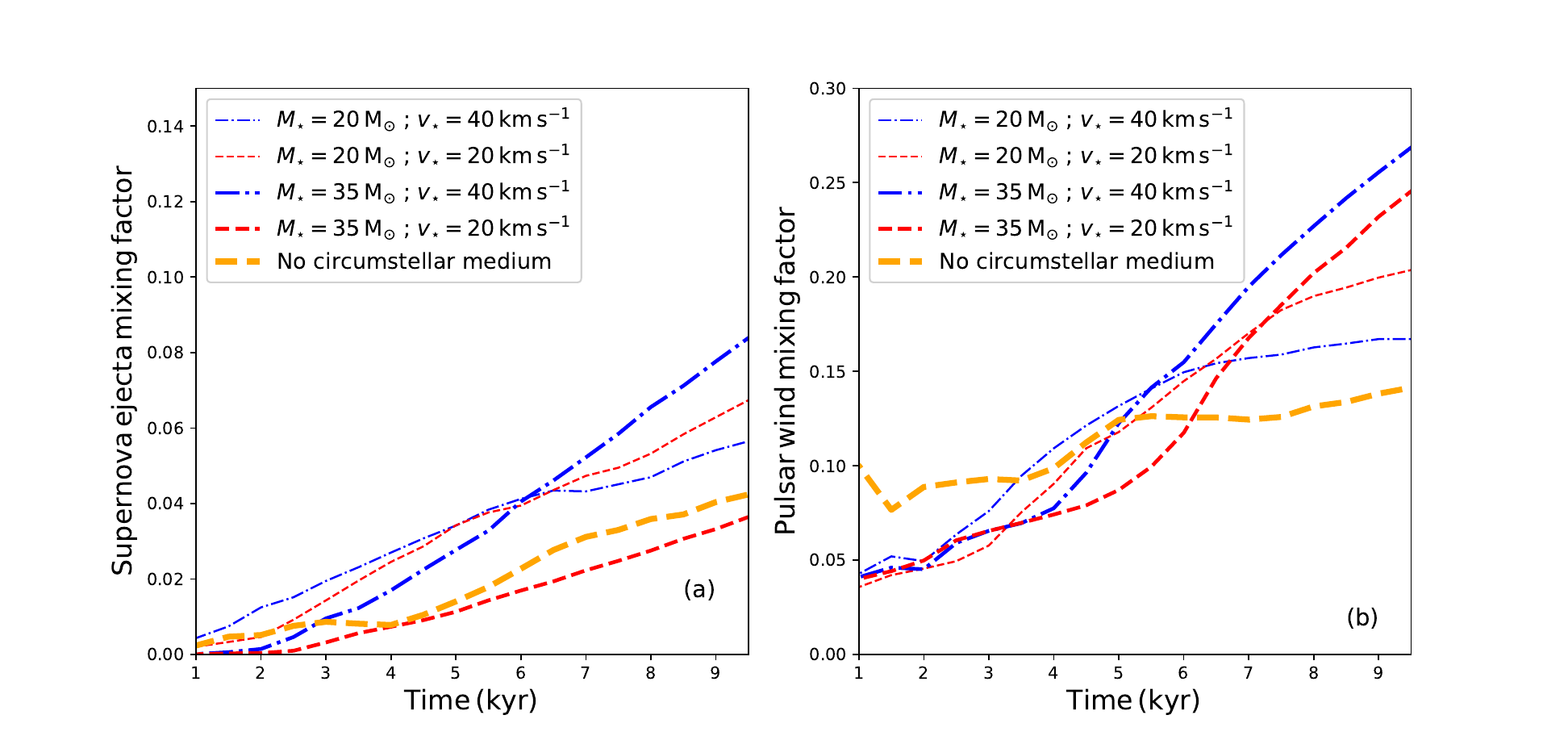}  \\      
        \caption{
        Mixing of the different materials in the plerionic supernova 
        remnant considered without circumstellar medium (orange line). 
        The line colors distinguish between the other 
        models with stellar velocity 
        $v_{\star}=20\, \rm km\, \rm s^{-1}$ (dashed red) and  
        $v_{\star}=40\, \rm km\, \rm s^{-1}$ (dotted dashed blue), 
        and for $M_{\star}=20\, \rm M_\odot$ (thin lines) and  
        $M_{\star}=35\, \rm M_\odot$ (thick lines).   
        The panels display the mixing for the supernova ejecta (a) 
        and pulsar wind (b), respectively. 
        }
        \label{fig:mixing_mhd_model_no_csm}  
\end{figure*}

The mixing efficiency of the supernova ejecta increase from $0\%$ to up to 
$8\%$ over the time that we consider in the calculations. This quantity increases 
quasi-linearly in the case of a $20\, \rm M_{\odot}$ progenitor until $6\, \rm kyr$
after the explosion, for both models with velocity $20\, \rm km\, \rm s^{-1}$ and 
$40\, \rm km\, \rm s^{-1}$, the later being more important than the sooner 
(Fig.~\ref{fig:mixing}c). 
The mixing in the models with a $35\, \rm M_{\odot}$ progenitor is milder until 
$6\, \rm kyr$, however, the faster model mixes better than the slowest model. 
Then, the simulation with velocity $40\, \rm km\, \rm s^{-1}$ rises quickly, 
reaching efficiency of $8\%$, as a direct consequence of the morphology of 
the circumstellar medium. 
The pulsar wind mixing efficiency evolves from $4\%$ to $15$$-$$20\%$ in the 
model with the progenitor $20\, \rm M_{\odot}$, while it reaches values as high 
as $25\%$ in the case of the $35\, \rm M_{\odot}$ star (Fig.~\ref{fig:mixing}d). 
At early times the mixing is more important in the red supergiant plerion, because 
the reverberation happens sooner since the pre-supernova bow shock is smaller. 
The pulsar wind recovers the X-shape in the model with $20\, \rm M_{\odot}$ and 
$40\, \rm km\, \rm s^{-1}$, which establishes an equilibrium between the region of 
pulwar wind and the region of ejecta, preventing further mixing. 
Inversely, mixing of pulsar wind material is very inefficient in the case of the 
$35\, \rm M_{\odot}$ progenitor that produced a very wide wind cavity prior to 
the explosion. The pulsar wind expands into the unperturbed ejecta until about 
$4\, \rm kyr$ after the explosion. Once the shock wave hits the cicumstellar 
medium and reflects, the pulsar wind is affected accordingly especially in the 
northern region of the supernova remnant, which increases greatly the mixing 
efficiency.

All in all, our work shows that the the mixing of material in the subsequent 
remnant is more important in the case of higher-mass runaway star moving at 
high velocities.  

The spatial distribution of the several tracers are further displayed at a 
selected time of the evolution of the pulsar wind nebulae in 
Fig. \ref{fig:cuts_tracers_20kms} (models with velocity $20\, \rm km\, \rm s^{-1}$) 
and in Fig. \ref{fig:cuts_tracers_40kms} (models with velocity $40\, \rm km\, \rm s^{-1}$), 
respectively. 
Horizontal (top panels) and vertical (bottom panels) cuts are taken through the supernova 
remnants generated by the supernova progenitor moving with $v_{\star}=20\, \rm km\, \rm s^{-1}$ 
prior to the explosion. The slices are along the R-direction direction (top figure) 
an along the z-direction (bottom figure). In each figure, the panel (a) concerns the 
$20\, \rm M_{\odot}$ progenitor star and the panel (b) concerns the $35\, \rm M_{\odot}$ 
progenitor star, respectively. The panels represent the number density 
(dashed purple line, in $\rm  cm^{-3}$), the main-sequence (cyan), red supergiant (orange), Wolf-Rayet (red), supernova ejecta (green) and pulsar wind (black) 
material tracers. 
The figures further illustrates that the most important mixing region is the 
location of the supernova remnants that are perpendicular to the direction 
of motion of the progenitor, where the blastwave meets the former stellar 
wind bow shock. It is clearly seen that the mixing of material is stronger 
in the case of a heavy progenitor star (see Fig. \ref{fig:cuts_tracers_20kms}a,b 
and Fig. \ref{fig:cuts_tracers_40kms}a,b), moving at high bulk velocity 
through the ISM prior to the explosion (see Fig. \ref{fig:cuts_tracers_20kms}a 
and Fig. \ref{fig:cuts_tracers_40kms}a, Fig. \ref{fig:cuts_tracers_20kms}b 
and Fig. \ref{fig:cuts_tracers_40kms}b). 

\section{Discussion}
\label{discussion}


\subsection{Limitations of the models}
\label{limitations}



\begin{figure*}
        \centering
        \includegraphics[width=1.7\columnwidth]{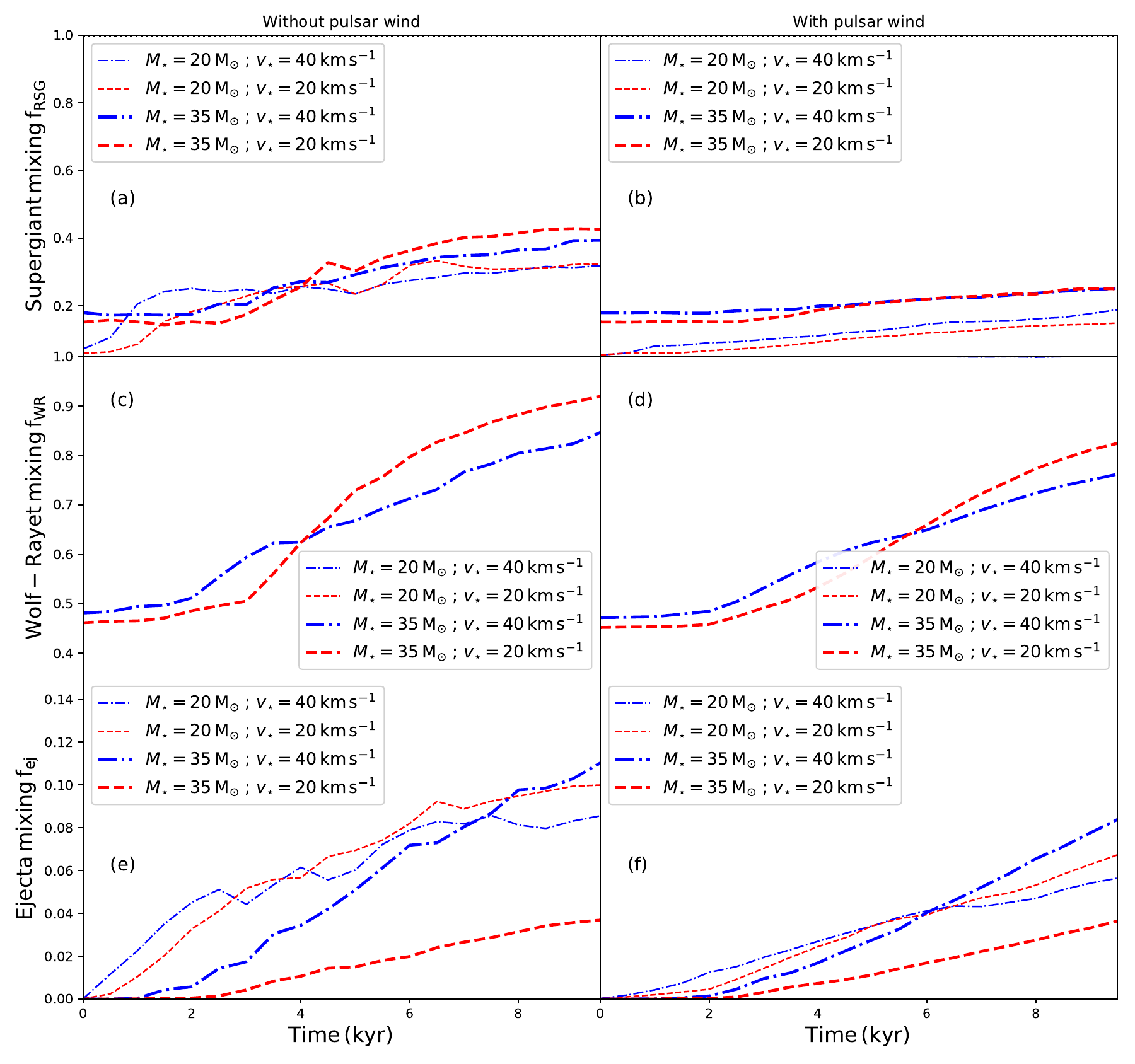}  \\      
        \caption{
        Mixing of the evolved stellar wind (top, middle panels) and supernova 
        ejecta (bottom panels) materials in the plerionic supernova remnants, 
        considered without (left) and with (right) pulsar wind. 
        The line colors distinguish between the other models with stellar velocity 
        $v_{\star}=20\, \rm km\, \rm s^{-1}$ (dashed red) and  
        $v_{\star}=40\, \rm km\, \rm s^{-1}$ (dotted dashed blue), 
        and for $M_{\star}=20\, \rm M_\odot$ (thin lines) and  
        $M_{\star}=35\, \rm M_\odot$ (thick lines).   
        }
        \label{fig:plot_comparison_mixing}  
\end{figure*}

\subsubsection{Axisymmetry, wind speed and magnetisation}

The 2.5-dimensional framework (with two spatial dimensions for the computational 
grid and three for the vectorial quantities) 
implies that any inclination of the supernova progenitor’s 
rotational axis with respect to its direction of motion is not treatable. 
Similarly, the pulsar’s 
spin axis in relation to the massive star’s movement and the local ISM magnetic field orientation 
are also not considered. 
Thus all the directions of the ISM magnetic field, progenitor's bulk motion, 
progenitor axis of rotation and pulsar axis of rotation are dictated by the symmetry axis of 
the cylindrical coordinate system \citep{chita_aa_488_2008,vanmarle_aa_541_2012,
chiotellis_mnras_502_2021}. 
Although this approach improves computational efficiency and yields useful insights into plerionic 
supernova remnants, it falls short of capturing the full complexity of these phenomena. 
Producing 
full 3-dimensional models \citep{herbst_apj_897_2020,velazquez_mnras_519_2023} would enable 
the inclusion of factors such as independent directions of motion and rotation of both 
the progenitor, respectively, as well as the pulsar bulk motion and update the works of 
\citet{kolb_apj_844_2017,temim_apj_851_2017,temim_apj_932_2022}.

\textcolor{black}{
The most prominent effect generated by the use of the above-described 2.5-dimensional 
coordinate system is the development of artificial jet-like features along the 
symmetry axis. These features appear when a source of momentum with a component 
parallel to the axis encounters a counterflow in its own reference frame. This 
situation arises in various cases, such as the bow shock of a massive runaway star 
\citep{comeron_aa_338_1998}, the wind bubble of a static OB star \citep{dwarkadas_apj_667_2007}, 
a protostellar jet \citep{mattia_aa_679_2023}, a supernova blast 
wave \citep{franco_pasp_103_1991,balsara_MNRAS_386_2008a}, 
or a pulsar wind nebula \citep{komissarov_mnras_349_2004}. 
This effect is further intensified by the presence of magnetic fields in the simulation 
\citep{meyer_mnras_464_2017}, by cooling mechanisms \citep{meyer_mnras_464_2017}, or by 
the radiative transport of ionizing emissions from the stellar object \citep{comeron_aa_326_1997}. 
In our present setup, combining a supersonic stellar wind, a runaway stellar 
object, a blast wave driven by a core-collapse supernova explosion, and the subsequent 
launch of a pulsar wind provides all the ingredients for the simulation models to 
develop such jet-like artifacts, as seen in 
Figs. \ref{fig:snr_pwn_2020}, \ref{fig:snr_pwn_2040}, \ref{fig:snr_pwn_3520}, 
\ref{fig:snr_pwn_3540}, and \ref{fig:plot_comparison_mixing}. As these features 
grow over time, they become more visible at times $\ge 10\, \rm kyr$.  
Discriminating between the artificial nature of these jet-like features and the 
development of physical structures, such as the polar jet of a pulsar wind nebula, 
would require full 3D simulations with high spatial resolution \citep{meyer_mnras_506_2021}, 
which is beyond the scope of this paper.
}

The pulsar wind nebulae that we examinate in our study are treated within the 
non-relativistic framework. By performing classical numerical magnetohydroddnamical 
simulations, we make use of a considerably slower pulsar wind speed that values 
corresponding to the high Lorentz factors that can be reached by such objects, 
see \citet{kennel_apj_283_1984}. 
A more stable simulation environment is achieved, which is better suited for long-term 
numerical investigations of supernova remnant and pulsar wind nebula problem. 
We are aware that a lower pulsar wind speed and/or its low magnetisation induce 
several drawbacks, particularly regarding the compression rates and shock speeds 
with the remnant and also the emergence of associated instabilities at the contact 
discontinuities in them. Our solution then applies to rather very weak-winded 
pulsars, however, they offer a first insight into the physics of material mixing 
in pulsar wind nebulae. 
Moreover, via the additional hypothesis that a substantial fraction of the magnetic 
field energy is transformed into kinetic energy in the pulsar wind nebulae, we 
assume in this study a low magnetization parameter of \(\sigma = 10^{-3}\), following 
the precedent works of \citet{kennel_apj_283_1984,2017hsn_book_2159S,
begelman_apj_397_1992}. 
We are aware that other numerical works showed that higher possible values  of 
\(\sigma \gg 0.01\) can be better suited for the explanation of the observational 
feature of some pulsar wind nebulae, see for example \citet{Porth_etal_2014MNRAS.438..278P}. 
This corner of the parameter space is still to be explored, the present study being 
more focused on the presentation of a method to study mixing in plerions, rather than 
on a systematic investigation of that question and/or on the tailoring of models to a 
specific object.  
%
%
This will be investigated in a forthcoming study.

\subsubsection{Other limitations}

Another class of improvement to be brought in our simulations concerns 
the initial conditions and the included microphysical processes. 
The pristine environment in which the progenitor high-mass stars move are 
still extremely simplified in the present models. 
Amongst others, the multi-phased nature of the ISM \citep{mckee_apj_195_1975} 
is to affect the development of pre-supernova wind nebulae such as the bow shocks that 
our runaway stars induce \citep{baalmann_aa_650_2021}.  
In turn, this will affect the subsequent blastwave-nebula interaction  
\citep{miceli_mnras_430_2013} and eventually the old, large-scale remnant 
and its emission properties \citep{das_aa_689_2024}. 
Including this element would provide a more accurate representation of the interaction between the 
pulsar wind and its surrounding environment. Moreover, considering the oblique rotating pulsar  
magnetosphere would enhance the accuracy of the simulated characteristics of pulsar wind nebulae, 
such as particle acceleration \citep{petri_mnras_512_2022} and non-thermal emission mechanisms 
see \citet{philippov_apJ_855_2018}. These improvements 
are essential for advancing our understanding and present propitious directions for future research.

Last, the current model investigate mixing of materials by use of passive scalars that are advected 
throughout the expanding supernova remnants and pulsar wind nebulae \citet{meyer_mnras_521_2023}, 
i.e. the time-dependent chemical evolution of the yields blown in the stellar winds and released 
in the ejecta are not included. 
This will request the coupling of the magneto-hydrodynamical 
models to a full chemical network \citet{grassi_mnras_439_2014,katz_mnras_512_2022} that change 
each species number density as a result of their natural interaction within the conditions of 
pressure and temperature in the remnant \citep{zhou_apj_93_2022}.  
Such models are highly desirable and we hope to present them in the future.

\subsection{The role of the circumstellar medium}
\label{csm}

Fig. \ref{fig:mhd_model_no_csm} displays the number density field in a magneto-hydrodynamical 
simulation of the supernova remnant of the $20\, \rm M_{\odot}$ progenitor star moving with 
velocity $v_{\star}=20\, \rm km\, \rm s^{-1}$, but excluding the influence of the wind-ISM 
interaction, i.e. the circumstellar nebula is not included into the calculation. The evolution 
of the plerionic supernova remnant is shown at times $15\, \rm kyr$ after the explosion and 
the two contours highlight the region with a $50\%$ contribution of pulsar wind (black) and 
supernova ejecta (white).  
The expansion of the forward shock of the blastwave is not constrained by the circumstellar 
medium and the overall supernova remnant adopts a circular morphology. Inside of it, the 
contact discontinuity between the pulsar wind and the supernova ejecta is affected by the 
Richtmeyer-Meshkov instability, a particular case of the Rayleight-Taylor instability 
\citep{kane_apj_511_1999}. The pulsar wind nebula produces the typical morphology that 
is described by the work of \citet{komissarov_mnras_349_2004}, with a diamond-like 
shape made of an equatorial ring and a vertical jet which propagation throughout the 
ejecta is magnified along the symmetry axis of the simulation domain. 
Note that a model of this kind has been presented in the work of 
\citet{meyer_mnras_515_2022} where it is shown that the presence of the 
circumstellar medium greatly affects the synchrotron radio emission 
signature of the pulsar wind nebulae.

In Fig. \ref{fig:mixing_mhd_model_no_csm} the mixing efficiency of the supernova ejecta 
and of the pulsar wind material of the plerion calculated excluding the circumstellar 
medium is plotted (thick dashed orange line) together with the other calculations 
including a runaway progenitor. 
The mixing efficiency of the pulsar wind nebula without wind-ISM interaction is lower 
than that of the other models, except for the simulation with a $35\, \rm M_{\odot}$ 
progenitor star moving with velocity $v_{\star}=20\, \rm km\, \rm s^{-1}$ (red dashed 
line of Fig. \ref{fig:mixing_mhd_model_no_csm}a). This is due to the very large cavity 
which is produced by the Wolf-Rayet stellar wind (Fig. \ref{fig:snr_pwn_3520}), 
permitting the blastwave to expand into a low-density medium during more time than in 
the other models before beginning to interact with the circumstellar medium. The mixing 
of the ejecta is therein slightly lower than when the blastwave is 
exploding into the warm phase of the ISM, which is denser than the Wolf-Rayet cavity 
in Fig. \ref{fig:snr_pwn_3520}. 
The evolution of the mixing efficiency of the pulsar wind material is initially larger 
than in the other simulations, however, when the supernova blastwave hits the stellar 
wind bow shock of the defunct progenitor and is reflected towards the center of the 
explosion, this translates into a stronger mixing than when the circumstellar medium 
is ignored (Fig. \ref{fig:mixing_mhd_model_no_csm}b). The presence of a complex medium 
at the moment of the explosion is therefore a factor that enhances mixing of material 
in plerionic supernova remnants.

\subsection{The role of the pulsar wind}
\label{role_pwn}

Fig.~\ref{fig:plot_comparison_mixing} displays the evolution of the mixing efficiency of the 
post-main-sequence stellar winds, i.e. red supergiant (a,b) and Wolf-Rayet (c,d), 
the supernova ejecta (e,f), considered without (left panels) and with (right panels). 
The quantities are plotted for the $20\, \rm M_{\odot}$ (thin lines) and the  
$35\, \rm M_{\odot}$ progenitor stars (thick lines), moving with velocities 
$v_{\star}=20\, \rm km\, \rm s^{-1}$ (dashed lines) and $v_{\star}=40\, \rm km\, \rm s^{-1}$ 
(dashed dotted lines).

To this end, additional numerical simulations without pulsar wind nebula are performed, 
using the same computational domains, initial conditions, boundary conditions and 
microphysical processes than in the models presented above. 
The absence of a pulsar wind following the supernova explosion mainly turns into the 
fact that the supernova blastwave, once interacting with the circumstellar medium 
of the progenitor, can freely reflect towards the center of the explosion, filling 
the interior of the supernova remnant 
\citep{meyer_mnras_450_2015,chiotellis_mnras_502_2021,meyer_mnras_521_2023}. 
As a consequence, the mixing efficiency is higher without pulsar wind (Fig.~\ref{fig:plot_comparison_mixing}a) 
than with it (Fig.~\ref{fig:plot_comparison_mixing}b), which reduces it by 
a factor of $2$ approximately. 
The same is true for the Wolf-Rayet stellar wind that reaches values to a mixing efficiency 
of $0.9$ to $0.8$ after $9\, \rm kyr$, see Fig.~\ref{fig:plot_comparison_mixing}c,d. 
Last, the mixing efficiency of the supernova ejecta are also affected by the 
absence of a pulsar wind, see the differences between its values in 
Fig.~\ref{fig:plot_comparison_mixing}e,f, which are lower in the plerionic case 
than in the case of supernova remnants without pulsar wind in them. 
We can conclude from this analysis that the presence of a pulsar wind nebula in a 
supernova remnant has a global effect consisting in diminishing the mixing of 
materials in it.

The detailed consequence of the pulsar wind blowing into the supernova remnant does not 
change the respective importance of the mixing efficiency of the red supergiant wind 
once the remnant is $7\, \rm kyr$ old, in the sense that this material in the models 
with a $35\, \rm M_{\odot}$ Wolf-Rayet progenitor star mix better than in the models 
with a red supergiant star (Fig.~\ref{fig:plot_comparison_mixing}a,b). 
The same is true for the Wolf-Rayet stellar wind, which is more important in the model 
with velocity $v_{\star}=20\, \rm km\, \rm s^{-1}$ than in the simulation with velocity 
$v_{\star}=40\, \rm km\, \rm s^{-1}$, see Fig.~\ref{fig:plot_comparison_mixing}c,d, 
and for the supernova ejecta, see Fig.~\ref{fig:plot_comparison_mixing}e,f. 
Most notable difference between the simulations with and without pulsar wind in them 
happen in the early evolution phase of the system, before that the blastwave hits the 
pre-supernova wind bubble and that the pulsar wind termination shock reverberates, 
at times $\leq 4$--$5\, \rm kyr$ after the explosion. Fluctuations in the time 
evolution of the mixing efficiency are particularly present in the models with a 
$20\, \rm M_{\odot}$ progenitor, as it produces a smaller bow shock prior to the 
explosion, provoking quicked reflection of the blaswtave and reverberation of the 
pulsar wind.

\subsection{\textcolor{black}{Mixing and emission properties of supernova 
remnants: towards possible future improvements}}
\label{emission}

\textcolor{black}{ 
We have shown that the stellar wind history of massive stars, in addition to 
shaping  their circumstellar medium prior to the explosion and governing the 
expansion of the supernova blastwave, also dictates the morphology of supernova remnants. 
This wind history is also responsible for the distribution of chemical elements 
within the remnants. 
The hydrogen-burning products from the main-sequence and 
red supergiant winds, the C, N, and O elements blown into the Wolf-Rayet material, 
and the 
heavy elements such as Mg, Si, Ca, Ti, and Fe in the supernova ejecta 
\citep{gabler_mnras_502_2021}, mix together through a series of shock-producing 
wave reflections and transmissions.
The final distribution of chemical yields is therefore the outcome of a complex 
process that begins at the onset of the main-sequence phase, coupling stellar 
evolution with the local conditions of the interstellar medium. Once the star 
has exploded, this process is powered by the pulsar wind from the magnetized 
neutron star formed during the core-collapse explosion and hosted by the remnant.
Understanding the precise locations of these chemical elements and their relative 
number densities enables us to determine the types of emissions they produce and 
the intensity of their fluxes. 
For example, main-sequence material emits H$\alpha$ 
at 6564 $\rm \mathring{A}$, evolved stellar winds 
produce atomic spectral lines such as the optical [N\,{\sc ii}] at 6584 
$\rm \mathring{A}$  and the [S\,{\sc ii}] doublet at 6716 and 6731 
$\rm \mathring{A}$. The forbidden [O\,{\sc iii}] line at 5007 $\rm \mathring{A}$ 
\citep{mavromatakis_aa_2003} and the O, Mg, and Fe elements generate soft X-ray 
emission in the keV energy band \citep{orlando_aa_622_2019}.
}

\textcolor{black}{
The approach we present here is primitive in that the tracers associated with each 
one of the species are passively advected with the plasma. 
In reality, chemical elements interact 
and undergo a variety of physical-chemical reactions, influenced by local temperature 
conditions. 
These reactions alter their abundances, resulting in the destruction of 
some atoms and the creation of new molecules, such as dust particles. The level of 
detail in the chemical reaction model directly impacts how closely the solution 
matches actual observations.
Several numerical methods have been developed to address this complexity. The PLUTO 
code, for instance, includes a limited network of chemical elements and reactions, 
allowing for non-equilibrium calculations of the ionization balance of various 
species. 
This approach is used to derive the appropriate gas cooling through 
collisionally excited line radiation \citep{mignone_apj_170_2007,
tesileanu_aa_488_2008,migmone_apjs_198_2012}. Similar methods, which also consider 
photoheating of gas by stellar ionizing radiation, have been presented by 
\citet{toala_apj_737_2011}. The KROME package allows users to define custom 
species and reactions, enabling the study of specific chemical phenomena when 
coupled with magneto-hydrodynamics \citep{grassi_mnras_439_2014}.
The distribution of chemical yields can be further analyzed using radiative transfer 
post-processing tools, which convert them into local emissivity maps, projected 
maps, and spectra.
}

\textcolor{black}{
Lastly, we take a closer look at the method derived by \citet{} in the study 
of historical supernova remnants, such as Cas A \citep{} and SN1987a \citep{}, 
which utilizes the same magneto-hydrodynamical code as ours.
High spatial resolution is achieved through a remapping routine that enables 
calculation of the supernova blastwave’s expansion into the stellar wind and 
circumstellar medium, while maintaining a constant number of grid zones. This 
approach provides an equivalent of the expanding grid used in codes such as ZEUS 
\citep{vanmarle_mnras_407_2010} for the PLUTO code. This mapping scheme includes 
tracers for ions such 
as H, He, C, O, Ne, Mg, Si, Ca, Ti, Ni, and Fe, provided by a self-consistent 
explosion simulation that incorporates neutrino physics, as shown by 
\citet{gabler_mnras_502_2021}.
By further accounting for line centroid shifts and broadening due to the Doppler 
effect, thermal broadening of emission lines, and uniform photoelectric absorption 
from the interstellar medium, one can, with the integration of a Boltzmann equation, 
obtain temperature and abundance profiles. These profiles allow synthesis of spectra 
using a non-equilibrium emission model, resulting in realistic X-ray spectra in the 
keV energy band \citep{orlando_aa_622_2019,orland_aa_636_2020,orlando_aa_666_2022,
2024arXiv240812462O}.
This procedure bridges the gap between simulations incorporating chemical reactions 
and synthetic observations, paving the way for a feedback loop of numerical 
experiments that converge on tailored models for specific astrophysical objects. 
Our research direction, focused on studying plerionic supernova remnants, will 
continue along these lines in the future.
}


\section{Conclusion}
\label{conclusion}

This research examines the influence of the 
circumstellar medium produced 
by a rotating, magnetized high-mass star that is exiled through the ISM exerts 
onto the mixing of the different kind of materials involved in the pulsar wind 
nebula developing inside of the supernova remnant formed after the explosive death 
of those runaway massive stars. 
The study presented aimed at 
determining the evolution of the 
spatial distribution of the progenitor's stellar winds, supernova ejecta, pulsar 
leptonic wind, and to quantify the manner they mix and melt into the remnant. 
The simulations were performed in the 2.5-dimensional magneto-hydrodynamical fashion, 
using the {\sc pluto} code \citep{mignone_apj_170_2007,migmone_apjs_198_2012}, with 
the microphysical processes and implementation described in \citet{meyer_mnras_521_2023}. 
The progenitors of the supernova remnants 
are considered to have formed into 
the warm phase of the Galactic plane of the Milky Way, to have been ejected 
and to move supersonically therein prior 
to finish their lives as a supernova. 

The considered progenitors are choosen to be the most common 
core-collapse supernova progenitors.
The selection takes into account their zero-age 
main-sequence mass ($20\, \rm M_\odot$ and $35\, \rm M_\odot$) according to the 
initial mass function \citep{kroupa_mnras_322_2001}, and bulk velocities 
($20\, \rm km\, \rm s^{-1}$ and $40\, \rm km\, \rm s^{-1}$) that bracket the 
peak of the velocity distribution of runaway OB stars 
\citep{blau1993ASPC...35..207B,renzo_aa_624_2019}, see also discussion regarding 
the initial conditions in \citet{meyer_aa_687_2024}. 
The pulsar we consider presents enegertics and timing features 
corresponding to that of \citet{2017hsn_book_2159S} and is to be taken as an example 
of the parameter space.

The simulation revel the following main results. 
\begin{itemize}
 \item The total amount of red supergiant wind mixes by $\le 20\%$, mostly due to the 
fact that large quantities of this material constitute the dense circumstellar 
medium and has filled the low-density trail of the runaway progenitor's bow shock, 
which are not yet impacted by the expansion of the supernova blastwave, regardless 
of the progenitor's mass and/or bulk velocity. 
\textcolor{black}{
This material is rich in H burning products which govern the 
H$\alpha$ emission contiunum of the supernova remnants.
} 
\item The distribution of Wolf-Rayet material is also strongly affected by the morphology 
of the circumstellar medium, which is rather circular if the $35\, \rm M_\odot$ 
progenitor star moves with velocity $20\, \rm km\, \rm s^{-1}$ as a result of 
the Wolf-Rayet shell forming prior to the explosion \citep{brighenti_mnras_273_1995} 
while this region is more oblong if the star move faster. The Wolf-Rayet wind 
mixes efficiently by $50\%$ to $80\%$ within the $10\, \rm  kyr$ after the explosion. 
\textcolor{black}{
The Wolf-Rayet material is enhanced in C, N, O elements, responsible 
for optical forbidden lines such as [N\,{\sc ii}] and [O\,{\sc iii}].
} 
\item The distribution of the supernova ejecta conserve sphericity for longer time 
if the progenitor moved with velocity $20\, \rm km\, \rm s^{-1}$, while it expands 
as complex patterns if the star moved faster, as a consequence of its earlier interaction 
with the circumstellar stellar wind bow shock. The ejecta mixing is more important 
for fast Wolf-Rayet progenitors ($8\%$) than for slower stars ($4\%$), whereas it is 
sensibly the same ($6\%$) in the context of a red supergiant progenitor.
\textcolor{black}{
The supernova ejecta, composed 
elements 
such as Mg, Si, Ca, Ti and Fe, 
induce soft keV emission and infrared photons, which 
can be observable. 
} 
\item Similarly, the pulsar wind mixes efficiently if the progenitor was Wolf-Rayet 
a star ($25\%$) than if it was a red supergiant star ($20\%$). 
\item The presence of a pulsar wind inside of a supernova remnant diminishes the 
mixing efficiency of the stellar wind and supernova ejecta in it. 
\end{itemize}

Our study shows that the past stellar and circumstellar evolution of massive stars 
govern the internal chemistry of their plerionic supernova remnants, which is a 
preponderant element to include in interpreting observations and deciphering the 
formation of these objects. 
%


\section*{Acknowledgements}

\textcolor{black}{
The author thankfully acknowledges RES resources provided by 
BSC in MareNostrum to AECT-2024-2-0002. 
The authors gratefully acknowledge the computing time made available to them 
on the high-performance computer 
"Lise" at the NHR Center NHR@ZIB. The latter 
center is jointly supported by the Federal Ministry of Education and 
Research and the state governments participating in the NHR 
(www.nhr-verein.de/unsere-partner). 
}
This work has been supported by the grant PID2021-124581OB-I00 funded by 
MCIU/AEI/10.13039/501100011033 and 2021SGR00426 of the Generalitat de Catalunya. 
This work was also supported by the Spanish program Unidad de Excelencia Mar\' ia 
de Maeztu CEX2020-001058-M and by the European Union 
NextGeneration program (PRTR-C17.I1).

\section*{Data availability}

This research made use of the {\sc pluto} code developed at the University of Torino  
by A.~Mignone (http://plutocode.ph.unito.it/). 
The figures have been produced using the Matplotlib plotting library for the 
Python programming language (https://matplotlib.org/). 
The data underlying this article will be shared on reasonable request to the 
corresponding author.


\bibliographystyle{mnras}

\bsp	
\label{lastpage}
\end{document}